\documentstyle[epsf]{mn}
\newif\ifAMStwofonts
\input psfig.sty

\newcommand{\mtot}{\relax \ifmmode M_{\rm tot}\else $M_{\rm tot}$\fi}
\newcommand{\Reff}{\relax \ifmmode R_{\rm e}\else $R_{\rm e}$\fi}
\newcommand{\SBe}{\relax \ifmmode \langle SB_{\rm e}\rangle \else $\langle SB_{\rm e}\rangle$\fi}
\newcommand{\mB}{\relax \ifmmode M_{\rm B}\else $M_{\rm B}$\fi}
\newcommand{\ReB}{\relax \ifmmode R_{\rm e,B}\else $R_{\rm e,B}$\fi}
\newcommand{\mueB}{\relax \ifmmode \mu_{\rm e,B} \else $\mu_{\rm e,B}$\fi}
\newcommand{\mD}{\relax \ifmmode M_{\rm D}\else $M_{\rm D}$\fi}
\newcommand{\muo}{\relax \ifmmode \mu_{\rm 0}\else $\mu_{\rm 0}$\fi}
\newcommand{\rd}{\relax \ifmmode h\else $h$\fi}
\newcommand{\db}{\relax \ifmmode D/B\else $D/B$\fi}
\newcommand{\nb}{\relax \ifmmode n\else $n$\fi}
\newcommand{\inc}{\relax \ifmmode i\else $h$\fi}
\newcommand{\ellip}{\relax \ifmmode \epsilon\else $\epsilon$\fi}

\newcommand{\magarc}{mag arcsec$^{-2}$}

\newcommand{\vmax}{\relax \ifmmode V_{\rm max}\else $V_{\rm max}$\fi}
\newcommand{\Rq}{$R^\frac{1}{4}$}
\newcommand{\Rn}{$R^\frac{1}{n}$}
\newcommand{\hi}{H {\sc I}}

\ifoldfss
  \ifCUPmtlplainloaded \else
    \NewTextAlphabet{textbfit} {cmbxti10} {}
    \NewTextAlphabet{textbfss} {cmssbx10} {}
    \NewMathAlphabet{mathbfit} {cmbxti10} {} % for math mode
    \NewMathAlphabet{mathbfss} {cmssbx10} {} %  "   "    "
  \fi
  \ifAMStwofonts
    \ifCUPmtlplainloaded \else
      \NewSymbolFont{upmath} {eurm10}
      \NewSymbolFont{AMSa} {msam10}
      \NewMathSymbol{\upi}     {0}{upmath}{19}
      \NewMathSymbol{\umu}     {0}{upmath}{16}
      \NewMathSymbol{\upartial}{0}{upmath}{40}
      \NewMathSymbol{\leqslant}{3}{AMSa}{36}
      \NewMathSymbol{\geqslant}{3}{AMSa}{3E}

      \let\leq=\leqslant 
       
    \fi
  \fi
\fi % End of OFSS

\ifnfssone
  \newmathalphabet{\mathit}
  \addtoversion{normal}{\mathit}{cmr}{m}{it}
  \addtoversion{bold}{\mathit}{cmr}{bx}{it}
  \newmathalphabet{\mathbfit} % math mode version of \textbfit{..}
  \addtoversion{normal}{\mathbfit}{cmr}{bx}{it}
  \addtoversion{bold}{\mathbfit}{cmr}{bx}{it}
  \newmathalphabet{\mathbfss} % math mode version of \textbfss{..}
  \addtoversion{normal}{\mathbfss}{cmss}{bx}{n}
  \addtoversion{bold}{\mathbfss}{cmss}{bx}{n}
  \ifAMStwofonts
    \ifCUPmtlplainloaded \else
      \UseAMStwoboldmath
      \makeatletter
      \new@mathgroup\upmath@group
      \define@mathgroup\mv@normal\upmath@group{eur}{m}{n}
      \define@mathgroup\mv@bold\upmath@group{eur}{b}{n}
      \edef\UPM{\hexnumber\upmath@group}
      \new@mathgroup\amsa@group
      \define@mathgroup\mv@normal\amsa@group{msa}{m}{n}
      \define@mathgroup\mv@bold\amsa@group{msa}{m}{n}
      \edef\AMSa{\hexnumber\amsa@group}
      \makeatother
      \mathchardef\upi="0\UPM19
      \mathchardef\umu="0\UPM16
      \mathchardef\upartial="0\UPM40
      \mathchardef\leqslant="3\AMSa36
      \mathchardef\geqslant="3\AMSa3E

      \let\leq=\leqslant 

    \fi
  \fi
\fi % End of NFSS release 1

\ifnfsstwo
  \DeclareMathAlphabet{\mathbfit}{OT1}{cmr}{bx}{it}
  \SetMathAlphabet\mathbfit{bold}{OT1}{cmr}{bx}{it}
  \DeclareMathAlphabet{\mathbfss}{OT1}{cmss}{bx}{n}
  \SetMathAlphabet\mathbfss{bold}{OT1}{cmss}{bx}{n}
  \ifAMStwofonts
    \ifCUPmtlplainloaded \else
      \DeclareSymbolFont{UPM}{U}{eur}{m}{n}
      \SetSymbolFont{UPM}{bold}{U}{eur}{b}{n}
      \DeclareSymbolFont{AMSa}{U}{msa}{m}{n}
      \DeclareMathSymbol{\upi}{0}{UPM}{"19}
      \DeclareMathSymbol{\umu}{0}{UPM}{"16}
      \DeclareMathSymbol{\upartial}{0}{UPM}{"40}
      \DeclareMathSymbol{\leqslant}{3}{AMSa}{"36}
      \DeclareMathSymbol{\geqslant}{3}{AMSa}{"3E}

      \let\leq=\leqslant 

    \fi
  \fi
\fi % End of NFSS release 2

\ifCUPmtlplainloaded \else
  \ifAMStwofonts \else % If no AMS fonts
    \def\upi{\pi}
    \def\umu{\mu}
    \def\upartial{\partial}
  \fi
\fi

\def\kms{\relax \ifmmode {\,\rm km\,s}^{-1}\else \,km\,s$^{-1}$\fi}
\def\ks{\relax \ifmmode  K_{\rm s}\else $K_{\rm s}$\fi}
\def\ha{\relax \ifmmode {\rm H}\alpha\else H$\alpha$\fi}
\def\hb{\relax \ifmmode {\rm H}\beta\else H$\beta$\fi}
\def\hi{\relax \ifmmode {\rm H\,{\sc i}}\else H\,{\sc i}\fi}
\def\hii{\relax \ifmmode {\rm H\,{\sc ii}}\else H\,{\sc ii}\fi}
\def\h2{\relax \ifmmode {\rm H}_2\else H$_2$\fi}
\def\lha{\relax \ifmmode L_{{\rm H}\alpha}\else $L_{{\rm H}\alpha}$\fi}
\def\shi{\relax \ifmmode \sigma_{{\rm HI}}\else $\sigma_{\rm HI}$\fi}
\def\sh2{\relax \ifmmode \sigma_{{\rm H}_2}\else $\sigma_{{\rm H}_2}$\fi}
\def\degr{\hbox{$^\circ$}}
\def\arcmin{\hbox{$^\prime$}}
\def\arcsec{\hbox{$^{\prime\prime}$}}
\def\deg{\hbox{$^\circ$}}
\def\min{\hbox{$^\prime$}}
\def\sec{\hbox{$^{\prime\prime}$}}
\def\fdg{\hbox{$.\!\!^\circ$}}
\def\fs{\hbox{$.\!\!^{\rm s}$}}
\def\farcm{\hbox{$.\mkern-4mu^\prime$}}
\def\farcs{\hbox{$.\!\!^{\prime\prime}$}}
\def\degd#1.#2{ #1\fdg#2 }                 % degrees over decimal point
\def\mind#1.#2{ #1\farcm#2 }               % minutes over decimal point
\def\secd#1.#2{ #1\farcs#2 }               % seconds over decimal point
\def\hhh{\ifmmode {\rm ^h}              % hours symbol
         \else {${\rm ^h}$}
         \fi}
\def\sss{\ifmmode {\rm ^s}              % seconds symbol
         \else {${\rm ^s}$}
         \fi}
\def\hms#1h#2m#3s{                      % hms format (for RA)
                  \relax
                  \ifmmode #1^{\rm h}\,#2^{\rm m}\,#3^{\rm s}
                  \else \hbox{$#1^{\rm h}\,#2^{\rm m}\,#3^{\rm s}$}
                  \fi
                 }
\def\dms#1d#2m#3s{                      % dms format (for Dec)
                  \relax
                  #1\degr\,#2\arcmin\,#3\arcsec 
                 }
\def\hmsd#1h#2m#3.#4s{                  % hms format with decimal point (RA)
                      \relax
                      \ifmmode #1^{\rm h}\,#2^{\rm m}\,#3\fs#4
                      \else \hbox{$#1^{\rm h}\,#2^{\rm m}\,#3\fs#4$}
                      \fi
                     }
\def\dmsd#1d#2m#3.#4s{                  % dms format with decimal point (Dec)
                      \relax
                      #1\degr\,#2\arcmin\,#3\farcs#4
                     }
\def\mag{\relax                          % magnitudes symbol
        \ifmmode ^{\rm m}
        \else $^{\rm m}$
        \fi
       }
\def\magd#1.#2{                          % magnitudes over decimal point
              \relax
              \ifmmode #1^{\rm m}
                       \hskip-0.55em.\hskip0.22em#2
              \else \hbox{#1$^{\rm m}
                    \hskip-0.55em.\hskip0.22em$#2}
              \fi
             }

\def\aj{AJ}                   % Astronomical Journal
\def\apj{ApJ}                 % Astrophysical Journal
\def\apjl{ApJ}                % Astrophysical Journal, Letters
\def\apjs{ApJS}               % Astrophysical Journal, Supplement
\def\apss{Ap\&SS}             % Astrophysics and Space Science
\def\aap{A\&A}                % Astronomy and Astrophysics
\def\aaps{A\&AS}              % Astronomy and Astrophysics, Supplement
\def\mnras{MNRAS}             % Monthly Notices of the RAS
\title{Structure and star formation in disk galaxies I. Sample selection 
and near infrared imaging}
\author[J.~H.~Knapen et al.]
       {J.~H.~Knapen$^{1,2}$, R.~S.~de~Jong$^3$, S.~Stedman$^1$ and
D.~M.~Bramich$^4$\\ 
        $^1$University of Hertfordshire, Department 
of Physical Sciences, Hatfield, Herts AL10 9AB\\
        $^2$Isaac Newton Group of Telescopes, Apartado 321, 
        E-38700 Santa Cruz de La Palma, Spain\\
        $^3$Space Telescope Science Institute, 3700 San Martin Drive, 
Baltimore, MD 21218, USA\\
	$^4$School of Physics and Astronomy, University of St. Andrews,
Scotland KY16 9SS}
\date{Accepted March 2003.
      Received ;
      in original form }

\pagerange{\pageref{firstpage}--\pageref{lastpage}}
\pubyear{2003}

\begin{document}

\maketitle

\label{firstpage}

\begin{abstract}

We present near-infrared imaging of a sample of 57 relatively large,
Northern spiral galaxies with low inclination. After describing the
selection criteria and some of the basic properties of the sample, we
give a detailed description of the data collection and reduction
procedures.  The \ks\ $\lambda=2.2\mu$m images cover most of the disk
for all galaxies, with a field of view of at least 4.2 arcmin. The
spatial resolution is better than an arcsec for most images. We fit
bulge and exponential disk components to radial profiles of the light
distribution. We then derive the basic parameters of these components,
as well as the bulge/disk ratio, and explore correlations of these
parameters with several galaxy parameters. 

\end{abstract}


\section{Introduction}

Near-infrared   (NIR)  imaging of galaxies is    a  better tracer of the
underlying stellar  structure than  optical   imaging, for two  reasons.
Firstly, the effects of   extinction by  dust   in the galaxy   are much
reduced (an order of  magnitude from $V$ to  $K$), and secondly, the NIR
emission  is a  better tracer  of the older  stellar  populations, which
contain most of the  stellar  mass. The combined   effect of these   two
factors  is that NIR emission  is a more accurate  tracer of the mass in
galaxies. This is also why the fraction of  galactic bars, mostly devoid
of current  star formation (SF), is  higher as determined  from NIR than
from optical surveys  ($\sim$70\%  versus $\sim55$\%, e.g., Sellwood  \&
Wilkinson 1993; Mulchaey \& Regan 1997; Knapen, Shlosman
\& Peletier 2000; Eskridge et al. 2000).

The main  technical problems  in NIR imaging  of spiral disks  are the
high background  at 2.2 micron, and  the relatively small  size of NIR
array detectors.  Detectors with 1024$\times$1024 pixels,  such as the
ones used  for the present  paper, have only  been in use  since about
1996,  and many  of the  cameras built  around such  arrays  have been
constructed  to exploit  the  higher spatial  resolution  in the  NIR,
and/or the  possibilities of adaptive  optics (e.g., KIR on  the CFHT,
Doyon et  al.  1998;  UFTI on  UKIRT, Roche et  al. 2002).  So whereas
optical  surveys of  galaxy  disks have  been  performed for  decades,
resulting, among many other results,  in galaxy catalogues such as the
RC3 (de Vaucouleurs et al.  1991), NIR surveys covering the whole disk
of nearby spiral galaxies are only now starting to be published (e.g.,
2MASS: Skrutskie  et al.  1997, Jarrett et  al. 2003; Seigar  \& James
1998a,  1998b; Moriondo  et  al.  1999;  M\"ollenhoff  \& Heidt  2001;
Eskridge et  al. 2002; this  paper). The NIR  cameras we used  for the
presently described project have been designed as a compromise between
sampling the  best seeing  conditions and a  relatively wide  field of
view  (INGRID on  the  4.2~m William  Herschel  Telescope, Packham  et
al. 2003), or specifically for wide-field imaging (PISCES on the 2.3~m
Bok telescope,  McCarthy et al. 2001). The  presently described survey
offers  several advantages over  most of  the other  studies mentioned
above, namely by  offering a set of high-resolution  (the median value
for the  FWHM seeing across  our sample is 0.90  arcsec), well-sampled
(pixel size 0.24 arcsec for  all but 4 images), reasonably deep images
of a  representative sample of  nearby, not highly  inclined, Northern
spiral galaxies.

Decomposition  of  galaxy  light  in  bulge and  disk  components  has
received renewed attention in recent years (e.g., Andredakis, Peletier
\& Balcells  1995; Byun \& Freeman  1995; de Jong  1996a; Graham 2001;
M\"ollenhoff \& Heidt 2001). One  of the main reasons for this renewed
attention stems from the realisation  that not all spheroids have \Rq\
de Vaucouleurs  (1948) type profiles,  but often are better  fitted by
\Rn\ Sersic  (1968) type profiles. Furthermore, it  has been suggested
that the shape of the bulge and its relation  to the disk may hold clues
about the formation history of the galaxy components (e.g., Courteau, de
Jong  \&   Broeils 1996;  Andredakis   1998; Graham  2001;   Trujillo et
al. 2002). It is now  generally believed that early-type spiral galaxies
have  bulges with de Vaucouleurs-like profiles  (i.e. Sersic \nb\ values
of order  4) and late-type spiral  galaxies bulges with exponential-like
profiles with $\nb\sim$1, probably with a  continuum in between.  It has
been suggested that the  exponential bulges in late-type spiral galaxies
are  the result of  bar instabilities driving  gas toward the centre and
scattering star in the vertical direction  by buckling resonances, while
the de Vaucouleurs  type bulges are the result  of accretion and merging
events, probably early in  the history of  the galaxy (e.g., Courteau et
al.  1996; Aguerri, Balcells \& Peletier 2001).  An alternative view was
put forward by Andredakis  et al. (1995),  who  proposed that the  outer
parts  of \Rq\ bulges  were strongly  affected by the  disk in late-type
spiral galaxies, resulting in bulges with lower \nb\ values.  Andredakis
(1998) found  that by growing a disk  adiabatically  on an existing \Rq\
bulge in a collisionless N-body simulation, he could indeed lower the
\nb\  value of  the  resulting bulge,  but  not to  values lower  than
$\nb=2$.

In this paper, we present  NIR imaging in the 2.2  micron \ks\ band of a
complete sample of  57  nearby,  not  highly inclined, Northern   spiral
galaxies of all types.  We also present fits  to the separate  bulge and
disk components   on the basis  of   radial surface brightness profiles.
Optical broad ($B, I$) and narrow (\ha)  band imaging of all galaxies in
our  sample will  be  presented in Paper~II  (Knapen   et al. 2003).  In
Paper~III (J.~H. Knapen, in preparation)  we study the morphology of the
\ha\  emission in the central few  kiloparsecs of the sample galaxies to
explore  the  relative frequencies  of,  for  instance, central peaks or
nuclear rings among the sample.  Correlations with morphological type or
nuclear non-stellar activity will then  be explored.  In further papers,
we will use ellipse fits to the NIR images  to explore the parameters of
the bars in  the  sample galaxies, and  will use  disk scale lengths  to
investigate whether, and in what fashion, these are affected by emission
from young  stars  in areas of  current  star formation.  The  currently
presented  data  set offers  many more  opportunities  for study, and as
examples we   mention the  study  by Block  et  al.  (2001) who  derived
gravitational torques  due to bars  from our NIR images,  and the one by
Jogee et al.  (2002),  who used one of  our NIR images, of  NGC~5248, to
argue  for  the  dynamical   connection  of  the  different   components
identified in that galaxy.

\section{The sample}

\subsection{Sample selection}

Our  sample of galaxies   was selected from the  list  of Elmegreen  and
Elmegreen (1987), which contains 708 galaxies  from the Second Reference
Catalogue  of Bright Galaxies (RC2;   de Vaucouleurs et al.  1976).  The
Elmegreen  and   Elmegreen sample  was  selected  to included  all those
galaxies listed   in   the  RC2   with    declination  $\delta>-35$\deg,
inclination     $i<60$\deg\    and   inclination-corrected  diameter  at
25\,\magarc, $D_{25}$, greater than 2 arcmin.  Our sample of 57 galaxies
was extracted from this large sample to give a smaller subset, primarily
for analysis of the spiral arm properties, as follows:

\begin{itemize}

\item All galaxies of 4.2 arcmin and above in diameter ($D_{25}$) were
selected to ensure that we could isolate the spiral arms, even at
lower spatial resolutions.

\item Galaxies with $\delta<-20$\deg\ were excluded, ensuring all
galaxies could be observed from the Northern hemisphere.

\item Galaxies of inclination higher than 50\deg, where we would have
difficulty isolating the spiral structure, were excluded.

\end{itemize}

This gave a sample of 57 galaxies  (Table\,\ref{sampletab}) of which six
have  a  diameter  larger than  10  arcmin.  The  limitations  of galaxy
catalogues are various,  due to their adopted  criteria of inclusion and
their  degree of completeness.  The RC2  from  which the original sample
was chosen is, as implied  by its title,  magnitude limited.  No attempt
has been made by us to investigate selection effects  in the sample, but
we note that the range  of galaxies included in our  sample is by nature
limited in apparent diameter, brightness, and thus also in redshift $z$.
This is a factor which  must be taken  into account with any conclusions
drawn from the data.

Our sample of  spiral galaxies covers the   full range of  morphological
types  from SA to SB,  the full range of  Hubble types, and the complete
range  of   arm    class from   1  to   12  (Elmegreen    and  Elmegreen
1987)\footnote{Elmegreen  \& Elmegreen (1987)    define arm classes   as
follows:  arm   class     1=chaotic, fragmented,   unsymmetrical   arms,
2=fragmented spiral  arm pieces with   no regular pattern,  3=fragmented
arms uniformly   distributed  around  the  galactic centre,   4=only one
prominent arm; otherwise fragmented arms, 5=two symmetric, short arms in
the   inner regions; irregular outer   arms, 6=two symmetric inner arms;
feathery  ringlike outer  structure, 7=two  symmetric,  long outer arms;
feathery or irregular inner arms, 8=tightly wrapped ringlike arms, 9=two
symmetric  inner arms; multiple  long   and continuous outer arms,   and
12=two  long symmetric  arms dominating the  optical  disk}.  The sample
contains normal and   active galaxies; isolated,  companion  and cluster
member    galaxies; disturbed   and   undisturbed  galaxies; barred  and
un-barred  galaxies;  galaxies with  and without  extensive massive star
forming regions;  and galaxies with and without  nuclear  rings or other
nuclear structure. Some of  the most important  of such basic parameters
of the sample galaxies are given  in Table\,\ref{sampletab}.  Stedman \&
Knapen (2001)  present    graphically the distribution   of   the sample
galaxies as a function of morphological type, arm  class, bar type, disk
diameter, ellipticity,  and systemic velocity, and  we thus refer to the
figures in that paper.

Of  the sample  galaxies,  23\% are  barred    (RC3 type SB),   26\% are
un-barred  (SA) and  51\% are  mixed  (SX), giving  a  total of 74\%  of
galaxies with  bar structure.  This  is consistent with the findings of,
e.g., Knapen et al. (2000), who estimate a bar fraction of approximately
2/3rds  for all galaxies.  Again  based on classifications from the RC3,
25\% of   the sample  galaxies  possess an   inner ring structure,  30\%
possess an inner  s-shaped structure  and  45\% are mixed.  The   sample
covers the full  range of Hubble  types from S0/a  to Sm,  with a marked
excess  of type bc  and c galaxies. In  fact, these two  classes of open
armed  spirals contribute  40\%   of  the total.   Every  arm  class  is
represented in the sample, with 45\% of the galaxies  in classes 7, 8, 9
and 12, denoting grand design spiral  arm structure. Our sample galaxies
are  fairly evenly  distributed   in  recession velocity, lying   mostly
between   750\,\kms and 1750\,\kms, with   a  few galaxies at velocities
between 2000 and 3000\,\kms.    We checked the  NASA/IPAC  Extragalactic
Database  (NED) for  nuclear  activity  in our   galaxies. As listed  in
Table\,\ref{sampletab}, 27 galaxies  (52\%)  are listed as  showing some
type  of nuclear activity,  of which almost  half  (13) are Seyferts and
most others LINERs.

We  conclude that although  the  sample  is   limited by its   selection
criteria, most importantly  a lower limit on  diameter  and is therefore
not  ``complete" in  a strictly   statistical  sense, it  is  internally
consistent and a fair, representative sample  of all galaxies within the
selection parameters, i.e., bright, relatively face-on spiral galaxies.

\begin{table*}
\hfill
%\centering
%begin{minipage}{200mm}
\begin{tabular}{lcccccccccccc}
\hline
\multicolumn{2}{c}{Galaxy}  & \multicolumn{2}{c}{Type} & Activity & $D_{25}$ & Arm & $i$ & PA & $M_B$ & $v$ & $D$ & Scale\\
NGC & M & (RC3) & (Carnegie) & (NED) & (\min) & Class &
(\deg) & (\deg) & & (km\,s$^{-1}$) & (Mpc) & (pc/\sec)\\
\hline 
0210 & & .SXS3 & Sb &  & 5 & 6 & 35 & 160 & 11.6 & 1634  & 20.3 & 98\\
0337A & & .SXS8 & Sc &  & 5.9 & 2 & 28 & 10 & 12.7 & 1076  & 13.7 & 66\\
0488 & & .SAR3 & Sab &  & 5.3 & 3 & 30 & 15 & 11.15 & 2269  & 29.3 & 142\\
0628 & 74 & .SAS5 & Sc &  & 10.5 & 9 & 16 & 25 & 9.95 & 656  & 9.7 & 47\\
0864 & & .SXT5 & Sbc &  & 4.7 & 5 & 28 & 20 & 11.4 & 1560  & 20.0 & 97\\
1042 & & .SXT6 & Sc &  & 4.7 & 9 & 27 & 15 & 11.56 & 1373  & 16.7 & 81\\
1068 & 77 & RSAT3 & Sb & Sy1 Sy2 & 7.1 & 3 & 22 & 70 & 9.61 & 1137  & 14.4 & 70\\
1073 & & .SBT5 & SBc &  & 4.9 & 5 & 16 & 15 & 11.47 & 1211  & 15.2 & 74\\
1169 & & .SXR3 & Sba &  & 4.2 & 3 & 34 & 28 & 12.2 & 2387  & 33.7 & 163\\
1179 & & .SXR6 & SBc &  & 4.9 & 3 & 27 & 35 & 12.6 & 1780  & 21.2 & 103\\
1300 & & .SBT4 & SBb/SB0 &  & 6.2 & 12 & 35 & 106 & 11.11 & 1568  & 18.8 & 91\\
2775 & & .SAR2 & Sa &  & 4.3 & 3 & 27 & 155 & 11.03 & 1354  & 17.0 & 82\\
2805 & & .SXT7 & &  & 6.3 & 5 & 28 & 125 & 11.52 & 1734  & 28.0 & 136\\
2985 & & PSAT2 & Sab & LINER & 4.6 & 3 & 26 & 0 & 11.18 & 1322  & 22.4 & 109\\
3184 & & .SXT6 & Sc & HII & 7.4 & 9 & 14 & 135 & 10.36 & 593  & 8.7 & 42\\
3227 & & .SXS1P & S0/Sb & Sy1.5 & 5.4 & 7 & 34 & 155 & 11.1 & 1157  & 20.6 & 100\\
3344 & & RSXR4 & Sbc &  & 7.1 & 9 & 16 & & 10.45 & 586  & 6.1 & 30\\
3351 & 95 & .SBR3 & SBb & HII Sbrst & 7.4 & 6 & 34 & 13 & 10.53 & 778  & 8.1 & 39\\
3368 & 96 & .SXT2 & Sab & Sy LINER & 7.6 & 8 & 33 & 5 & 10.11 & 897  & 8.1 & 39\\
3486 & & .SXR5 & Sbc & Sy2 & 7.1 & 9 & 30 & 80 & 11.05 & 682  & 7.4 & 36\\
3631 & & .SAS5 & Sbc &  & 5 & 9 & 11 & & 11.01 & 1158  & 21.6 & 105\\
3726 & & .SXR5 & Sbc &  & 6.2 & 5 & 33 & 10 & 10.91 & 849  & 17.0 & 82\\
3810 & & .SAT5 & Sc &  & 4.3 & 2 & 32 & 15 & 11.35 & 994  & 16.9 & 82\\
4030 & & .SAS4 & &  & 4.2 & 9 & 31 & 27 & 12.02 & 1460  & 25.9 & 126\\
4051 & & .SXT4 & Sbc & Sy1.5 & 5.3 & 5 & 30 & 135 & 10.83 & 725  & 17.0 & 82\\
4123 & & .SBR5 & SBbc & Sbrst HII & 4.4 & 9 & 30 & 135 & 11.98 & 1329  & 25.3 & 123\\
4145 & & .SXT7 & Sc & HII/LINER & 5.9 & 4 & 31 & 100 & 11.78 & 1016  & 20.7 & 100\\
4151 & & PSXT2* & Sab & Sy1.5 & 6.3 & 5 & 32 & 50 & 11.5 & 995  & 20.3 & 98\\
4242 & & .SXS8 & Sd/SBd &  & 5 & 1 & 28 & 25 & 11.37 & 517  & 7.5 & 36\\
4254 & 99 & .SAS5 & Sc &  & 5.4 & 9 & 20 & & 10.44 & 2407  & 16.8 & 81\\
4303 & 61 & .SXT4 & Sc & HII Sy2 & 6.5 & 9 & 14 & & 10.18 & 1569  & 15.2 & 74\\
4314 & & .SBT1 & SBa & LINER & 4.2 & 12 & 18 & & 11.43 & 963  & 9.7 & 47\\
4321 & 100 & .SXS4 & Sc & LINER HII & 7.4 & 12 & 22 & 30 & 10.05 & 1586  & 16.8 & 81\\
4395 & & .SAS9* & Sd/SBd & LINER Sy1.8 & 13.2 & 1 & 23 & 147 & 10.64 & 320  & 3.6 & 18\\
4450 & & .SAS2 & Sab & LINER & 5.3 & 12 & 30 & 175 & 10.9 & 1956  & 16.8 & 81\\
4487 & & .SXT6 & Sc &  & 4.2 & 5 & 34 & 100 & 12.26 & 1037  & 19.9 & 97\\
4535 & & .SXS5 & SBc &  & 7.1 & 9 & 32 & 0 & 10.59 & 1957  & 16.8 & 81\\
4548 & 91 & .SBT3 & SBb & LINER Sy & 5.4 & 5 & 26 & 150 & 10.96 & 486  & 16.8 & 81\\
4579 & 58 & .SXT3 & Sab & LINER Sy1.9 & 5.9 & 9 & 26 & 95 & 10.48 & 1519  & 16.8 & 81\\
4618 & & .SBT9 & SBbc & HII & 4.2 & 4 & 24 & 25 & 11.22 & 544  & 7.3 & 35\\
4689 & & .SAT4 & Sc &  & 4.3 & 3 & 24 & & 11.6 & 1619  & 16.8 & 81\\
4725 & & .SXR2P & SBb & Sy2 & 10.7 & 6 & 32 & 35 & 10.11 & 1206  & 12.4 & 60\\
4736 & 94 & RSAR2 & Sab & Sy2 LINER & 11.2 & 3 & 24 & 105 & 8.99 & 310  & 4.3 & 21\\
5247 & & .SAS4 & Sc &  & 5.6 & 9 & 20 & 20 & 10.5 & 1357  & 22.2 & 108\\
5248 & & .SXT4 & Sbc & Sy2 HII & 6.2 & 12 & 31 & 110 & 10.97 & 1153  & 22.7 & 110\\
5334 & & .SBT5* & SBc &  & 4.2 & 2 & 31 & 15 & 11.99 & 1382  & 24.7 & 120\\
5371 & & .SXT4 & Sb/SBb & LINER & 4.2 & 9 & 26 & 8 & 11.32 & 2553  & 37.8 & 183\\
5457 & 101 & .SXT6 & Sc &  & 28.8 & 9 & 14 & & 8.31 & 241  & 5.4 & 26\\
5474 & & .SAS6P & Scd/SBcd & HII & 4.8 & 2 & 18 & & 11.28 & 273  & 6.0 & 29\\
5850 & & .SBR3 & SBb &  & 4.3 & 8 & 20 & 140 & 11.54 & 2556  & 28.5 & 138\\
5921 & & .SBR4 & SBbc & LINER & 4.9 & 8 & 24 & 130 & 11.49 & 1480  & 25.2 & 122\\
5964 & & .SBT7 & &  & 4.2 & 2 & 27 & 145 & 12.6 & 1447  & 24.7 & 120\\
6140 & & .SBS6P & &  & 6.3 & 2 & 31 & 95 & 11.81 & 910  & 18.6 & 90\\
6384 & & .SXR4 & Sb & LINER & 6.2 & 9 & 35 & 30 & 11.14 & 1663  & 26.6 & 129\\
6946 & & .SXT6 & Sc & HII & 11.5 & 9 & 22 & & 9.61 & 52  & 5.5 & 27\\
7727 & & SXS1P & Sa &  & 4.7 & 1 & 28 & 35 & 11.5 & 1814  & 23.3 & 113\\
7741 & & .SBS6 & SBc &  & 4.4 & 5 & 34 & 170 & 11.84 & 755  & 12.3 & 60\\
\hline
\end{tabular}

\caption{Global  parameters of  the galaxies  in the  observed sample.
Except   where  otherwise   mentioned,  all   figures  are   from  the
RC3. Tabulated  are NGC (column~1)  and Messier numbers  (column~2) of
the sample  galaxies; morphological type  from the RC3  (column~3) and
from the Carnegie Atlas of Galaxies (Sandage \& Bedke 1994; column~4);
nuclear  activity from  the NED  (column~5); apparent  major isophotal
diameter  measured   at  or   reduced  to  surface   brightness  level
$\mu_B=25.0$\,$B$-\magarc  ($D_{25}$; column~6);  Elmegreen  arm class
(from  Elmegreen and Elmegreen  1987; column~7),  where the  arm class
ranges  from 1 (flocculent)  to 12  (grand design);  inclination ($i$;
column~8),  from  the  ratio  of  the major  to  the  minor  isophotal
diameter; position angle PA of  the major axis of the disk (column~9);
total magnitude  $M_B$ (column~10); mean  heliocentric radial velocity
as  derived  from  neutral  hydrogen  observations  ($v$;  column~11);
distance from the Nearby  Galaxies Catalog (Tully 1988; column~12) and
image scale (column~13), as derived from the distance.}

\label{sampletab}
\end{table*}

\section{Observations and data reduction}

\subsection{INGRID observations}

Images of all but four of our sample  galaxies were obtained through the
\ks\ filter  with the  INGRID NIR  camera (Packham et   al. 2003) at the
Cassegrain focus of  the 4.2~m  William Herschel  Telescope  (WHT).  The
INGRID camera  is built around  a 1024$\times1024$ pixel Hawaii detector
array     (HgCaTe),  and gives   a      projected pixel  size  of  0.242
arcsec\footnote{The pixel size of  INGRID is  0.238  arcsec as  of March
22nd, 2001.  The few images we obtained  after this date were re-gridded
to the original  pixel size of 0.242  arcsec.} with a  field of  view of
$4.2\times4.2$ arcmin.  Observations for this programme were made during
a total of  six allocated observing nights  (May 14-16, 2000, Nov.  8 \&
12, 2000   and Dec. 7, 2000),   while additional images   were collected
during a number of service nights (Nov. 3, Dec.  4,  2000; March 9, Oct.
4, 2001;  and Jan. 4, 2002).  The  image of NGC~3351 was obtained during
the commissioning of INGRID, on the night of March 18, 2000.  Conditions
during most  of the nights were good,  photometric and with good seeing.
Whereas for some  galaxies images could only be  obtained  on one night,
many were observed during different nights.  The images of worst quality
in terms of  sky subtraction are of a  number of galaxies around RA=14h,
which  could  only be observed  during   our May  2000 nights, when  the
observations were hampered by clouds.  Total exposure times range from 6
to 100 minutes on source, as  listed in Table~\,\ref{imagetab}.  Typical
1$\sigma$ background  noise is 20.5~mag\,arcsec$^{-2}$, depending on the
observing  time,  down   to  21.7~mag\,arcsec$^{-2}$   for  our  longest
exposure, of NGC~7741. This compares favourably with the value given for
2MASS images of  bright galaxies, of 20.0~mag\,arcsec$^{-2}$ (Jarrett et
al. 2003).

The   observing  strategy   consisted  of   small  blocks   of  galaxy
observations,  each  of some  5  minutes  duration, interspersed  with
similar blocks of exposures of a nearby area of background sky, for an
equal amount  of time. The blocks  consisted of 4  pointings, where in
each pointing  the galaxy centre was  displaced by about  10 arcsec on
the array. Each  of these pointings in  turn was built up from  3 to 5
individual exposures of  20 to 12 seconds, co-averaged  to produce one
FITS file after 60 second  total exposure time. The high background in
the \ks\ filter,  variable and mostly dependent on  the temperature of
the outside air and the telescope,  set the limit to the length of the
individual  exposures. The  so-called  dither pattern  of 4  pointings
serves to  eliminate the  bad pixels during  the reduction,  while the
background  sky  exposures  are  paramount for  sky  subtraction  (see
below).

Whenever conditions were photometric,  standard stars from the list of
Hawarden et al.  (2001) were observed at regular  intervals during the
night.

\subsection{INGRID data reduction}

The data reduction  was performed using a combination  of standard and
purpose-built IRAF and  IDL tasks. The main reduction  script has been
adapted from a  set of IDL procedures originally  written by F. Rigaut
and  R. Doyon  to  reduce images  taken  with the  KIR  camera on  the
CFHT. The first step is the creation of a background frame from offset
sky exposures taken before and  after the galaxy frames. The different
background frames  are combined using an  iterative median combination
algorithm in order  to remove stellar images and  bad pixels which may
be  present in  the individual  exposures.  In each  of the  (normally
three)  iterations  the  individual   frames  are  compared  with  the
median-combined frame produced from them, and any pixels which lead to
peaks or  troughs in the combined  image of more than  three times the
noise are set  to undefined in the offending single  frame in the next
iteration.

The background  frame is then  subtracted from each  individual galaxy
frame, and the background-subtracted galaxy frames are median-combined
into the  reduced image.  In this latter  step, bad pixels  are masked
out,  and flatfielding  is done  on the  individual frames.  Since the
galaxy is at  a slightly different position in  each individual frame,
the position of either  the nucleus of the galaxy or of  a star in the
field  is determined  and used  to shift  the images  before combining
them.

Where several images  of the same galaxy were  available, either taken
during  the same  night  or  on different  nights,  these images  were
averaged after  registering them  using the positions  of a  number of
foreground  stars, and/or the  nucleus of  the galaxy.  The procedures
used  for  the alignment  were  very  similar  to those  described  in
$\S$3.4, below, for the registering of NIR and optical images. What we
will refer to as ``final'' images are those which have been registered
with the  optical images of  our data set  (as described in  detail in
Paper~II).

Photometric  calibration was  performed  by comparison  of the  final,
combined,  images with the  reduced images  as obtained  on individual
photometric  nights.  For  those  nights, calibration  constants  were
derived  from  the  observations  of photometric  standard  stars.  We
measured  the total  flux in  the two  images in  a  circular aperture
centred on  the nucleus  of the galaxy,  making sure that  exactly the
same area  was taken for  both the photometrically calibrated  and the
final image.  Correcting  for background and pixel size,  which are in
some  cases different  in the  photometric and  final images,  we thus
calibrate the final images.

\subsection{PISCES observations and data reduction}

Images of four  of our sample galaxies were  obtained using the PISCES
camera on the Bok 2.3~m telescope of the Steward Observatory, on Oct.\
17 (NGC\,6946) and 18 (NGC\,628,  1068, 6184), 1999. The PISCES camera
(McCarthy et  al. 2001)  uses the  same kind of  array as  INGRID, but
gives pixels of \secd 0.5 on the sky. The details of the observations,
data  reduction  and  calibration  procedures  for  these  images  are
described  in McCarthy  et  al.  (2001) and  only  the essentials  are
repeated here. The  PISCES camera has an 8.5  arcmin circular field of
view and the galaxies were mosaiced with a few central images and many
overlapping  images in the  outer parts,  interspaced with  sky offset
images.  Flatfields were created by differencing a number of early and
late  twilight   images.  The  flatfielded  object   images  were  sky
subtracted  by  weight-averaged and  clipped  offset  sky frames.  The
object images were corrected  for image distortions and then optimally
combined  using cross-correlations  for positional  offsets  and using
airmass  coefficient calibrations  for  intensity matching.  Effective
exposure times range typically from  about 1000~s in the outer regions
to 2000~s in  the centre. Seeing on the combined  images is 2.5 arcsec
for  the  Oct.\ 17  observation  and 1.5  arcsec  for  those of  Oct.\
18. Calibration was  performed by observing standard  star fields from
the list  of Hunt et al.  (1998) at least  4 times a night  at several
airmasses, to allow accurate calculation of airmass and colour terms.

\subsection{Final data sets}

\begin{table}
\hfill
\centering
%begin{minipage}{200mm}
\begin{tabular}{lcccc}
\hline
Galaxy &  Telescope &  Exp. time & Seeing & Size\\
 & & (min) & (arcsec) & (arcmin)\\
\hline
NGC 0210 & WHT & 24 & 1.04 & 4.2\\
NGC 0337A & WHT & 28 & 1.04 & 4.2\\
NGC 0488 & WHT & 45 & 1.68 & 4.9\\
NGC 0628 & Bok & N/A & 1.55 & 11.0\\
NGC 0864 & WHT & 24 & 0.94 & 4.2\\
NGC 1042 & WHT & 45 & 2.16 & 4.8\\
NGC 1068 & Bok & N/A & 1.64 & 9.6\\
NGC 1073 & WHT & 24 & 1.02 & 4.2\\
NGC 1169 & WHT & 24 & 0.78 & 4.2\\
NGC 1179 & WHT & 28 & 0.89 & 4.2\\
NGC 1300 & WHT & 32 & 1.03 & 5.2\\
NGC 2775 & WHT & 12 & 0.78 & 4.2\\
NGC 2805 & WHT & 48 & 1.06 & 4.2\\
NGC 2985 & WHT & 20 & 1.14 & 4.2\\
NGC 3184 & WHT & 24 & 0.89 & 5.1\\
NGC 3227 & WHT & 12 & 0.73 & 4.7\\
NGC 3344 & WHT & 24 & 0.85 & 4.9\\
NGC 3351 & WHT & 12 & 1.14 & 7.4\\
NGC 3368 & WHT & 8 & 0.77 & 5.0\\
NGC 3486 & WHT & 24 & 0.83 & 4.9\\
NGC 3631 & WHT & 38 & 0.87 & 4.2\\
NGC 3726 & WHT & 12 & 0.69 & 4.2\\
NGC 3810 & WHT & 12 & 0.69 & 4.2\\
NGC 4030 & WHT & 30 & 0.69 & 4.2\\
NGC 4051 & WHT & 12 & 0.74 & 4.8\\
NGC 4123 & WHT & 28 & 0.76 & 4.2\\
NGC 4145 & WHT & 36 & 1.04 & 4.2\\
NGC 4151 & WHT & 12 & 0.75 & 4.8\\
NGC 4242 & WHT & 24 & 1.26 & 4.2\\
NGC 4254 & WHT & 12 & 0.82 & 4.2\\
NGC 4303 & WHT & 19 & 1.06 & 4.8\\
NGC 4314 & WHT & 35 & 2.02 & 4.2\\
NGC 4321 & WHT & 12 & 0.76 & 5.0\\
NGC 4395 & WHT & 20 & 1.04 & 4.8\\
NGC 4450 & WHT & 12 & 0.89 & 4.7\\
NGC 4487 & WHT & 13 & 0.83 & 4.2\\
NGC 4535 & WHT & 12 & 0.77 & 5.0\\
NGC 4548 & WHT & 12 & 0.81 & 4.8\\
NGC 4579 & WHT & 12 & 0.98 & 4.8\\
NGC 4618 & WHT & 28 & 1.16 & 4.2\\
NGC 4689 & WHT & 12 & 0.81 & 4.2\\
NGC 4725 & WHT & 24 & 0.97 & 6.0\\
NGC 4736 & WHT & 19 & 0.98 & 5.2\\
NGC 5247 & WHT & 12 & 0.81 & 4.8\\
NGC 5248 & WHT & 12 & 0.79 & 4.8\\
NGC 5334 & WHT & 16 & 0.94 & 4.2\\
NGC 5371 & WHT & 12 & 0.91 & 4.2\\
NGC 5457 & WHT & 6 & 3.03 & 4.2\\
NGC 5474 & WHT & 16 & 1.00 & 4.2\\
NGC 5850 & WHT & 12 & 0.93 & 4.2\\
NGC 5921 & WHT & 12 & 1.02 & 4.2\\
NGC 5964 & WHT & 32 & 0.82 & 4.2\\
NGC 6140 & Bok & N/A & 1.74 & 8.0\\
NGC 6384 & WHT & 8 & 0.70 & 4.2\\
NGC 6946 & Bok & N/A & 2.31 & 11.6\\
NGC 7727 & WHT & 56 & 0.82 & 4.2\\
NGC 7741 & WHT & 75 & 0.90 & 4.8\\
\hline
\end{tabular}

\caption{Properties of the final  NIR images: telescope used to obtain
the image (col.~2), total on-source exposure time in minutes (col.~3),
seeing  as   measured   from  the   final   image  (in   arcsec;
col.~4), and field of view on the final image (in arcmin, where images 
are assumed to be square; col.~5).}

\label{imagetab}
\end{table}

Since we are  interested in comparing the NIR  images with the optical
images  we have  obtained  for our  sample  galaxies, we  have used  a
special procedure  to make the  NIR images directly comparable  to the
others. This  procedure is described  in detail in Paper~II,  where we
also  describe the optical  observations and  their reduction,  but in
outline the following  was done. We determined the  positions of three
foreground stars (or failing that,  two stars plus the galaxy nucleus)
in each of the images to be combined, as well as in an astrometrically
calibrated  image which  we  obtained from  the  Digitized Sky  Survey
(DSS). From  these positions, an  IRAF script we  developed calculates
the translation, rotation, and scaling  needed for each image. It also
determines which  image has  the highest resolution  (smallest pixels)
and  the largest  extent. Finally,  the rotation  needed to  place the
image set in an RA, dec orientation is determined from the position of
the stars in  the DSS image.  Images are  then rotated, translated and
scaled in such a way that the resulting images are correctly oriented,
have the  pixel size of  the image with  the smallest pixels,  and the
extend of  that covering  the largest area.   Those images  covering a
smaller  area are  filled with  zero-valued pixels.   Even  though the
resulting  images may  contain many  more  pixels, and  are thus  much
larger,  this  is  the only  way  to  ensure  that no  information  is
lost. Table\,\ref{imagetab}  gives the size of the  fully reduced \ks\
images,  but before  the operations  described immediately  above. The
area given  in the Table  thus represents the  useful size of  the NIR
images.   Table\,\ref{imagetab} also  lists with  which  telescope the
images were obtained, as well as the total on-source exposure time and
the spatial  resolution. The NIR images  of all galaxies  are shown in
Fig.\,\ref{images}.

\begin{figure*}
\caption{$K_{\rm s}$ images of all galaxies in the sample. North is up
and East to the left in all images, and the scale of each individual
image is indicated by the white bar in the bottom right corner, which
represents a length of 1 arcmin. Indicative contours are shown at surface 
brightness levels of 17 and 15\,\magarc\ for all galaxies, {\it except}
the following, for which contours are shown at 16 and 15\,\magarc:
NGC~2775, NGC~3368, NGC~4030, NGC~4051, NGC~4254, NGC~4303, NGC~4535,
NGC~4548, NGC~4736, NGC~5248, NGC~6384.}
\label{images}
\end{figure*}

%\setcounter{figure}{0}
%\begin{figure*}
%\psfig{figure=panel2.ps,width=19cm}
%\caption{Continued.}
%\label{images}
%\end{figure*}

%\setcounter{figure}{0}
%\begin{figure*}
%\psfig{figure=panel3.ps,width=19cm}
%\caption{Continued.}
%\label{images}
%\end{figure*}

\section{Bulge/disk decomposition}

In recent  years, performing  bulge/disk   (B/D) decompositions in   two
dimensions, i.e., on the full image, instead of in one dimension, on the
luminosity profiles, has become the accepted norm (e.g., Byun \& Freeman
1995; de  Jong 1996a; M\"ollenhoff \&   Heidt 2001; Simard  et al. 2002;
Tran et al. 2003). We have made 2D fits to our $K$-band images using the
GIM2D  code (Simard et al. 2002).   Unfortunately, due to limitations in
the  signal-to-noise  ratio  and image   extent,  and due   to mosaicing
imperfections, 2D fitting  of our $K$-band images   proved to be  highly
unstable. We therefore resorted to 1D B/D fitting,  pointing out that at
least the resulting disk parameters  (and most of the bulge  parameters)
will be equal within  the errors of  the fit in  the 1D and 2D cases (de
Jong 1996a;  MacArthur et al.   2003). In addition, for face-on galaxies
1D and 2D fitting is nearly  identical when no non-axisymmetric features
(bars, arms) are fitted and error weighting is done consistently.

We used the  IRAF ellipse fitting tasks to fit  radial profiles to the
surface brightness distribution  in all images. We used  sets of fixed
position angle ellipticity for each  galaxy, as taken from the RC3 and
checked  with our NIR  and optical  (Paper~II) images.   The resulting
azimuthally averaged profiles are shown in Fig.~\ref{profilefig}.  The
1D fitting technique  we used is very similar to  the one described in
de Jong (1996a), with some modifications reflecting the investigations
described  by  MacArthur  et  al.   (2003).   The  Levenberg-Marquardt
nonlinear least-squares  fitting technique was used. For  the disks we
used  exponential profiles,  for the  bulges Sersic  profiles.  Before
fitting,  the  model  profiles  were convolved  with  Gaussian  seeing
profiles with FWHMs equal to the FWHMs measured on foreground stars in
the  images. To  reduce the  effects  of uncertainties  in the  seeing
profiles, we discarded  the inner few points, namely  those with radii
less than two tenths of the FWHM seeing value. Relative weights of the
profile   data  points   were  set   to  reflect   photon  statistics,
uncertainties due to sky  measurement errors, seeing errors and errors
due  to  unmodelled  features.   The latter  correction  is  necessary
because  the light profiles  contain features  that are  not modelled,
like bars and spiral arms, which cause residuals in the fit (see e.g.,
de Jong 1996a, his fig.\,3).  Not taking the unmatched components into
account will result in reduced $\chi^2$ values much larger than 1.  We
have used  an uncertainty of  0.05 mag rms  at each data point  due to
unmatched components in our weighting scheme.

As shown  by MacArthur et al.  (2003), fitting Sersic  profiles with a
free \nb\  value is  very unstable and  often results  in non-physical
results. We have used a similar grid search technique as them, fitting
the profiles with fixed Sersic \nb\ values, with \nb\ ranging from 0.5
to 5 in steps of 0.1. Unlike MacArthur et al. (2003), we did not use a
combination of an inner and a  global $\chi^2$ to select the best fit,
but used only the $\chi^2$ of the whole profile fit to select the best
fit.

All fits were  repeated with the maximum sky  error estimate added and
subtracted  from the  luminosity  profile.  This  provides a  reliable
estimate of the uncertainties in  the bulge and disk parameters, as it
has shown before that sky errors dominate B/D decompositions (e.g., de
Jong 1996a;  MacArthur et  al.~2003).  The fits  of many  galaxies had
large  errors because of  the limited  field of  view of  the detector
compared  to  the galaxy  size  and the  limits  of  mosaicing in  the
near-IR. We  therefore decided to  exclude all disk parameters  if the
uncertainty in  their central surface brightness (\muo)  or disk scale
length (\rd)  was larger than 30\%  or if the fitted  scale length was
larger than half the image size. Likewise, bulge fits were rejected if
the error in the effective  radius (\ReB) or in the surface brightness
at this radius (\mueB) was larger than 30\%, if \ReB\ was smaller than
the  FWHM of  the seeing,  or if  $\nb\leq 0.8$  (as is  the  case for
NGC\,3351 and  NGC\,7741 which  have disturbed centres).   This strict
selection leaves  only 13 of the  57 galaxies with  both reliable disk
and bulge  parameters. In Fig.\,\ref{profilefig} we show  the fits for
all those galaxies where either bulge or disk fit was reliable.

\begin{table*}
\begin{tabular}{lrrrrrrr}
\hline
Name & \multicolumn{1}{c}{\mueB} & \multicolumn{1}{c}{\ReB} & \multicolumn{1}{c}{\nb} & \multicolumn{1}{c}{\muo} & \multicolumn{1}{c}{\rd}\\
 & \multicolumn{1}{c}{(\magarc)} & \multicolumn{1}{c}{(arcsec)} & & \multicolumn{1}{c}{(\magarc)} & \multicolumn{1}{c}{(arcsec)} \\
\hline
NGC\,0337A & $21.08^{+0.08}_{-0.08}$ & $ 25.73^{+  5.49}_{-  7.42}$ & $0.9^{+0.1}_{-0.1}$ & $19.79^{+0.30}_{-0.32}$ & $ 65.40^{+ 20.32}_{- 38.61}$\\[1mm]
NGC\,0488 & $16.35^{+0.13}_{-0.37}$ & $  9.20^{+  0.87}_{-  2.65}$ & $1.8^{+0.1}_{-0.3}$ & $16.64^{+0.15}_{-0.30}$ & $ 34.18^{+  3.81}_{-  7.40}$\\[1mm]
NGC\,0628 & $18.24^{+0.03}_{-0.14}$ & $ 22.44^{+  0.80}_{-  3.20}$ & $1.5^{+0.1}_{-0.1}$ & $17.68^{+0.05}_{-0.12}$ & $ 82.42^{+  5.86}_{-  9.40}$\\[1mm]
NGC\,0864 & $18.36^{+0.34}_{-1.09}$ & $  6.81^{+  1.37}_{-  6.74}$ & $3.0^{+0.3}_{-1.0}$ & $17.47^{+0.09}_{-0.18}$ & $ 31.86^{+  3.20}_{-  5.22}$\\[1mm]
NGC\,1042 & $18.64^{+0.21}_{-0.40}$ & $  6.42^{+  0.78}_{-  1.73}$ & $2.1^{+0.2}_{-0.4}$ & $18.69^{+0.07}_{-0.10}$ & $ 49.40^{+  6.88}_{-  9.23}$\\[1mm]
NGC\,1169 & $16.42^{+0.12}_{-0.13}$ & $  7.22^{+  0.59}_{-  0.71}$ & $1.6^{+0.1}_{-0.1}$ & $16.81^{+0.10}_{-0.12}$ & $ 33.49^{+  2.97}_{-  3.68}$\\[1mm]
NGC\,1179 & $19.31^{+0.07}_{-0.14}$ & $  5.54^{+  0.34}_{-  0.68}$ & $1.7^{+0.1}_{-0.1}$ & $18.92^{+0.04}_{-0.08}$ & $ 31.57^{+  2.33}_{-  5.01}$\\[1mm]
NGC\,1300 & $16.56^{+0.03}_{-0.11}$ & $  7.24^{+  0.17}_{-  0.48}$ & $1.9^{+0.1}_{-0.1}$ & $18.05^{+0.06}_{-0.10}$ & $ 48.71^{+  2.39}_{-  9.73}$\\[1mm]
NGC\,2775 & $16.34^{+0.26}_{-0.28}$ & $  9.27^{+  1.55}_{-  2.12}$ & $2.4^{+0.2}_{-0.2}$ & $15.95^{+0.15}_{-0.19}$ & $ 26.20^{+  2.28}_{-  3.17}$\\[1mm]
NGC\,3227 & $14.36^{+0.11}_{-0.20}$ & $  2.86^{+  0.16}_{-  0.32}$ & $1.9^{+0.1}_{-0.2}$ & $16.27^{+0.12}_{-0.17}$ & $ 25.64^{+  3.11}_{-  4.74}$\\[1mm]
NGC\,3351 & $15.45^{+0.03}_{-0.03}$ & $  9.46^{+  0.18}_{-  0.18}$ & $0.8^{+0.1}_{-0.1}$ & $16.58^{+0.06}_{-0.05}$ & $ 45.44^{+  5.30}_{-  5.48}$\\[1mm]
NGC\,3631 & $17.21^{+0.11}_{-0.03}$ & $ 10.30^{+  0.69}_{-  0.27}$ & $1.8^{+0.1}_{-0.1}$ & $18.65^{+0.16}_{-0.08}$ & $ 64.58^{+  7.11}_{-  5.32}$\\[1mm]
NGC\,3810 & $16.15^{+0.25}_{-0.13}$ & $  3.43^{+  0.59}_{-  0.37}$ & $1.5^{+0.2}_{-0.1}$ & $15.68^{+0.12}_{-0.06}$ & $ 15.96^{+  1.70}_{-  0.91}$\\[1mm]
NGC\,4051 & $14.85^{+0.18}_{-0.19}$ & $  3.22^{+  0.29}_{-  0.34}$ & $2.5^{+0.2}_{-0.2}$ & $17.24^{+0.15}_{-0.14}$ & $ 48.23^{+ 11.97}_{- 17.52}$\\[1mm]
NGC\,4303 & $15.03^{+0.09}_{-0.03}$ & $  4.46^{+  0.19}_{-  0.09}$ & $1.3^{+0.1}_{-0.1}$ & $16.51^{+0.06}_{-0.04}$ & $ 39.54^{+  3.72}_{-  3.68}$\\[1mm]
NGC\,4535 & $16.19^{+0.21}_{-0.11}$ & $  3.98^{+  0.42}_{-  0.27}$ & $2.1^{+0.2}_{-0.1}$ & $17.57^{+0.07}_{-0.04}$ & $ 63.45^{+ 11.07}_{- 11.28}$\\[1mm]
NGC\,4689 & $19.10^{+0.14}_{-0.31}$ & $ 14.79^{+  1.83}_{-  5.07}$ & $2.1^{+0.1}_{-0.2}$ & $17.78^{+0.07}_{-0.18}$ & $ 42.95^{+  5.58}_{- 11.15}$\\[1mm]
NGC\,4725 & $16.46^{+0.13}_{-0.13}$ & $ 16.16^{+  1.35}_{-  1.55}$ & $2.5^{+0.1}_{-0.1}$ & $17.30^{+0.14}_{-0.14}$ & $ 83.67^{+ 15.25}_{- 21.72}$\\[1mm]
NGC\,5247 & $18.88^{+0.05}_{-0.15}$ & $ 23.38^{+  1.58}_{-  3.03}$ & $2.3^{+0.1}_{-0.1}$ & $18.22^{+0.13}_{-0.30}$ & $ 90.97^{+ 16.96}_{- 31.96}$\\[1mm]
NGC\,5371 & $16.20^{+0.03}_{-0.03}$ & $  6.06^{+  0.17}_{-  0.17}$ & $1.8^{+0.1}_{-0.1}$ & $17.14^{+0.05}_{-0.05}$ & $ 51.44^{+  6.52}_{- 10.73}$\\[1mm]
NGC\,5457 & $18.63^{+0.14}_{-0.06}$ & $ 21.82^{+  2.23}_{-  1.64}$ & $2.1^{+0.1}_{-0.1}$ & $17.50^{+0.06}_{-0.06}$ & $128.07^{+ 17.13}_{- 14.97}$\\[1mm]
NGC\,5474 & $19.46^{+0.03}_{-0.03}$ & $ 18.16^{+  0.75}_{-  0.75}$ & $0.9^{+0.1}_{-0.1}$ & $19.71^{+0.13}_{-0.13}$ & $ 81.29^{+ 24.87}_{- 30.05}$\\[1mm]
NGC\,5850 & $17.50^{+0.14}_{-0.25}$ & $ 14.05^{+  1.46}_{-  2.68}$ & $2.9^{+0.1}_{-0.2}$ & $18.25^{+0.33}_{-0.41}$ & $ 39.11^{+ 10.05}_{- 19.20}$\\[1mm]
NGC\,5921 & $15.54^{+0.12}_{-0.22}$ & $  3.59^{+  0.27}_{-  0.50}$ & $1.6^{+0.1}_{-0.2}$ & $17.37^{+0.23}_{-0.26}$ & $ 30.43^{+ 10.74}_{- 23.89}$\\[1mm]
NGC\,6140 & $19.79^{+0.06}_{-0.08}$ & $ 22.74^{+  4.66}_{-  6.34}$ & $1.1^{+0.1}_{-0.1}$ & $19.03^{+0.40}_{-0.59}$ & $ 43.45^{+  8.51}_{- 17.11}$\\[1mm]
NGC\,6384 & $16.81^{+0.27}_{-0.15}$ & $  9.99^{+  1.83}_{-  1.43}$ & $1.9^{+0.2}_{-0.1}$ & $16.96^{+0.24}_{-0.20}$ & $ 41.09^{+  9.27}_{- 12.77}$\\[1mm]
NGC\,6946 & $16.09^{+0.22}_{-0.04}$ & $ 12.24^{+  1.47}_{-  0.30}$ & $2.2^{+0.2}_{-0.1}$ & $16.95^{+0.10}_{-0.02}$ & $127.18^{+ 21.50}_{-  2.28}$\\[1mm]
NGC\,7727 & $16.18^{+0.14}_{-0.24}$ & $  9.14^{+  0.87}_{-  1.56}$ & $2.4^{+0.1}_{-0.2}$ & $17.28^{+0.29}_{-0.39}$ & $ 29.58^{+  5.32}_{-  8.90}$\\[1mm]
NGC\,7741 & $18.88^{+0.04}_{-0.04}$ & $  8.43^{+  0.36}_{-  0.36}$ & $0.6^{+0.1}_{-0.1}$ & $18.38^{+0.05}_{-0.07}$ & $ 38.76^{+  1.93}_{-  8.14}$\\[1mm]
\hline
\end{tabular}
\caption{Fitted bulge and disk parameters, not corrected for extinction
or inclination}
\label{fitpartab}
\end{table*}

Fitting results for those galaxies with at least a reliable bulge or a
reliable   disk   fit   are   presented  in   Fig.\,\ref{typenb}   and
Table\,\ref{fitpartab}. Fits are not  plotted for those galaxies where
both bulge and disk parameters could not reliably be determined. Bulge
or disk parameters that were poorly determined are marked by a star in
Fig.~\ref{profilefig}.

\setcounter{figure}{1}
\begin{figure}
\psfig{figure=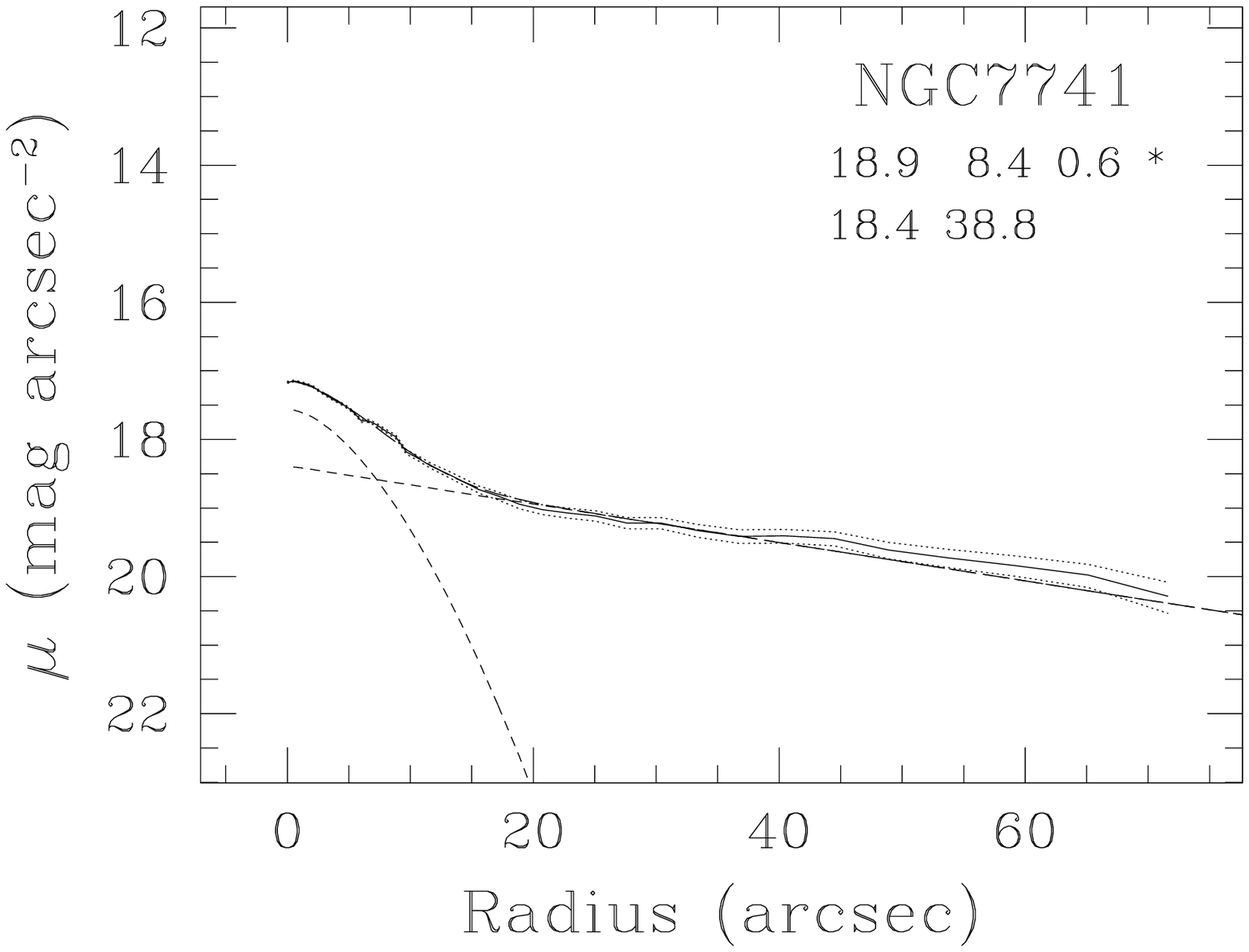,width=9cm}
\caption{Continued.}
\label{profilefig}
\end{figure}

\setcounter{figure}{2}
\begin{figure}
\epsfxsize=8.2cm
\epsfbox[54 147 300 717]{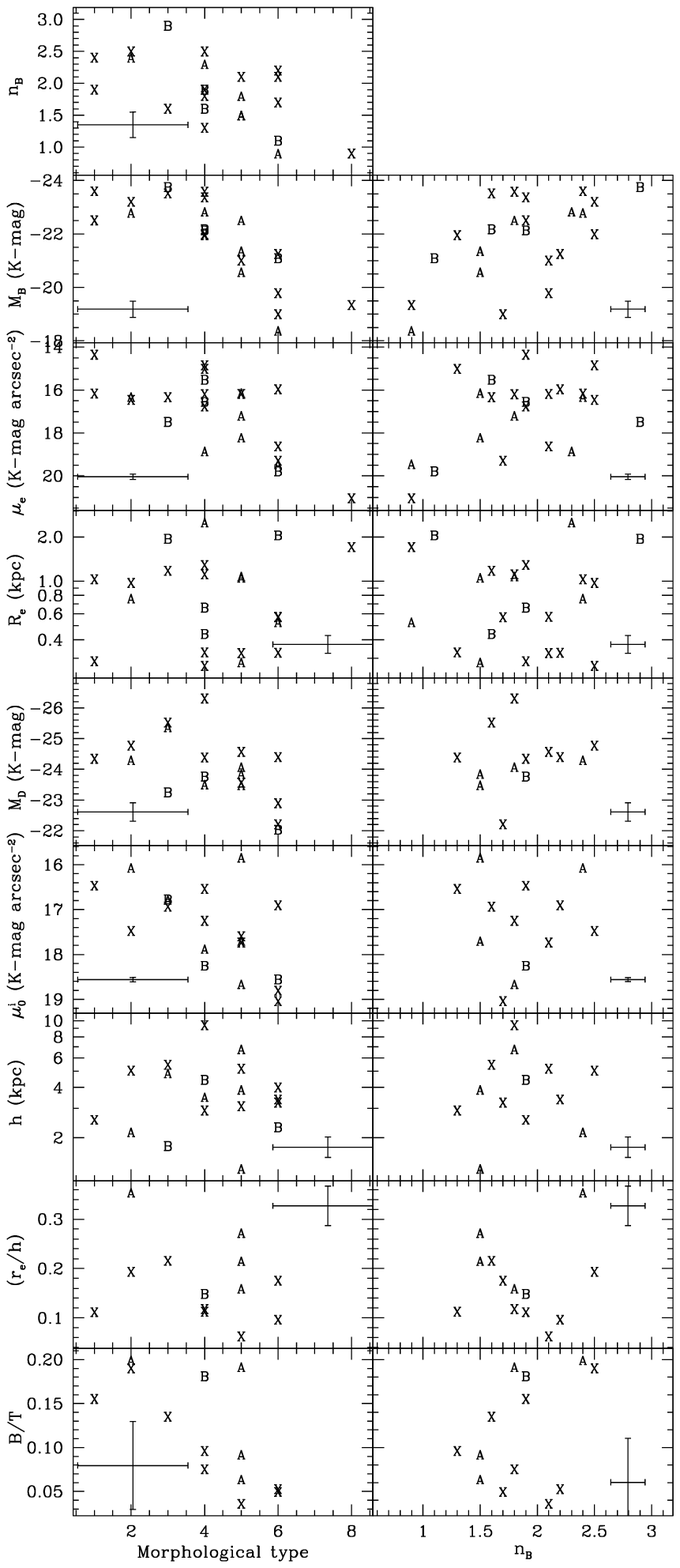}
\caption{
Various disk and bulge parameters as function of morphological T-type
and Sersic \nb\ bulge shape parameter. Non-barred galaxies are
indicated by A, barred galaxies by B and X are used for intermediate
types. The errorbars indicate the median errors for the plotted galaxies.
}
\label{typenb}
\end{figure}

In what  follows,  photometric  parameters were corrected   for Galactic
foreground extinction following the  precepts of Schlegel, Finkbeiner \&
Davis  (1998).  Distances to put measurements  at an absolute scale were
obtained from the  Nearby  Galaxies Catalog (Tully 1988).  Disk  surface
brightnesses  were  corrected   to  face-on  values   using the    fully
transparent correction of $\muo^i  = \muo-2.5\log(1-\ellip)$,  where the
ellipticities  $\ellip$ were taken from the  RC3 but double-checked with
the outer parts of the images.

Many combinations between  the various bulge  and disk parameters can be
investigated, but   with  so few reliable   decompositions   we refer to
previous   work for most of      them (e.g., Moriondo    et al.    1998;
M\"ollenhoff  \& Heidt  2001; Graham  2001). We note,  however, that for
spiral galaxies  correlations between the  disk parameters  are severely
affected by selection effects, most  notably in the case of correlations
involving \muo\ and  \rd.  Intrinsically large, high  surface brightness
galaxies are simply more likely to enter a magnitude or diameter limited
sample than small, low surface  brightness galaxies (Disney \& Phillipps
1983; Allen \& Shu 1979).  These selection effects have to be taken into
account  to calculate   proper distributions (de   Jong   \& Lacey 2000;
Blanton    et al.~2001; Cross \&   Driver  2002).  Furthermore, if bulge
parameters are tightly    correlated  with disk parameters  (like    the
$h/\ReB$ ratio  discussed  below), then  also the  distribution of bulge
parameters will be biased.

In Fig.\,\ref{typenb}  we show the reliable bulge  and disk parameters
as function of morphological T-type (from the RC3) and of best fitting
\nb.  As noted  before (e.g.,  de  Jong 1996b;  M\"ollenhoff \&  Heidt
2001), the change from T-type ~2 to ~8 is strongly correlated with the
change in bulge luminosity. The decrease in luminosity with increasing
T-type is mainly the result of bulge surface brightness changes helped
a little  bit by bulge  shape changes, but  is not due to  bulge scale
size changes.

Figure\,\ref{typenb}  also shows  that the  disk luminosities  peak at
about T-type  3-4, decreasing on average for  earlier and later-types,
driven  mainly by  surface brightness  but  partly by  scale size,  as
observed  before  by  de   Jong  (1996b)  and  M\"ollenhoff  \&  Heidt
(2001). The  relative bulge to disk  scale size ratio  does not change
much in this morphological type  range as often observed before (e.g.,
de Jong  1996b; Courteau  et al. 1996;  Moriondo et al.   1998; Graham
2001; MacArthur et  al.  2003). This implies that  the change observed
in $B/T$ ratio as a function  of morphological type, a change which is
in itself not surprising as the $B/T$ ratio is one of the criteria for
defining  morphological type,  is  mainly due  to  differences in  the
relative effective surface brightnesses of the bulge and the disk.

\begin{table}
\begin{tabular}{lcc}
\hline
Parameters & $r_{\rm s}$ & Significance\\
\hline
Type -- $B/T$ & -0.75 & 0.047\\
Type -- \mueB & 0.46 & 0.169\\
Type -- $M_B$ & 0.81 & 0.002\\
\nb\ -- $M_B$ & -0.54 & 0.083\\
Type -- \nb & -0.53 & 0.090\\
\hline
\end{tabular}
\caption{For a number of selected correlations, as discussed in the text
(column~1), we list the Spearman rank-order correlation coefficient
$r_{\rm s}$ (column~2) and significance (column~3), i.e., the
probability that this relation occurs at random.}
\label{corrs}
\end{table}

When now  turning to  the bulge and  disk parameter  correlations with
\nb,  we  confirm  previously  known  trends between  \nb\  and  bulge
luminosity and surface    brightness   (e.g., Andredakis et    al  1995;
M\"ollenhoff \& Heidt 2001; Trujillo et al.   2002).  These authors also
report  other correlations between   \nb\ and, for  instance,  \ReB\ and
$B/T$, but we have too few data points to confirm such trends.  Finally,
there is  a clear trend between morphological  type and \nb, observed by
most other authors, though not as strongly  by MacArthur et al.  (2003).
We present correlation  coefficients and significance indicators of some
of the  trends mentioned in Table\,\ref{corrs}.   Almost all  trends and
parameter ranges  are consistent with  continuations of trends seen in a
large sample of early-type galaxies studied by R.~S.~de  Jong et al. (in
preparation).

\section{Concluding remarks}

In this paper, we present the first set of results of our study of the
spiral arm  and disk properties of  a sample of  57 nearby, relatively
face-on,  spiral  galaxies.   We  describe the  sample  selection  and
summarise some of the properties of the sample, and present a full set
of NIR \ks-band images of  the sample galaxies. Those images have been
obtained  using the 4.2~m  WHT and  2.3~m Bok  telescopes, and  with a
minimum image size of 4.2~arcmin square  cover most of the disk of all
galaxies. We  describe the  data collection and  reduction procedures,
and show all resulting images. 

We  derive radial  profiles to  the light  distribution of  all sample
galaxies,   and  perform  1D   bulge/disk  decomposition   by  fitting
exponential  disk and  Sersic  bulge profiles.   The  results of  this
exercise are  shown, but due  to the limited signal-to-noise  ratio in
our images we can fit both bulge and disk parameters in a reliable way
for only 13 of our 57 galaxies, although either reliable bulge or disk
parameters  can be determined  for more  galaxies. Using  the reliable
bulge   and  disk   parameters  thus   derived,  we   explore  several
correlations   between   those  and   galactic   parameters  such   as
morphological  type. We  confirm correlations  between bulge  and disk
luminosities and morphological type,  and between the Sersic parameter
$n$  and bulge  luminosity and  surface brightness,  and morphological
type. 

In further  papers in  this series, we  will publish  the accompanying
optical  broad ($B,  I$) and  narrow-band  (\ha) imaging,  as well  as
analyses  of disk scale  lengths, and  of properties  of morphological
entities such as bars and rings within the galaxies.

\section*{Acknowledgments}

JHK thanks Mr.  J.~Knapen  for his assistance  and moral  support during
the observations  on Nov.  12, 2000.   We thank Dr.   D.~W. McCarthy for
making his PISCES camera available for this survey and Dr.  McCarthy and
R.~A.  Finn for their generous support during observations. We thank Dr.
C.  Packham for his assistance during the  early INGRID observations and
for sharing his INGRID expertise  with us. Drs.  F.  Rigaut and R. Doyon
kindly made their NIR image reduction software available  to us. The WHT
is operated on the island of La  Palma by the  Isaac Newton Group in the
Spanish  Observatorio  del Roque de   los Muchachos of  the Instituto de
Astrof\'\i  sica de Canarias.  The  Digitized Sky Survey was produced at
the  Space Telescope Science Institute  under   US Government grant  NAG
W-2166.This  research    has made use    of  the NASA/IPAC Extragalactic
Database  (NED) which is    operated by the  Jet Propulsion  Laboratory,
California  Institute of Technology,   under contract with  the National
Aeronautics and Space Administration.

\setcounter{figure}{1}
\begin{figure*}
\psfig{figure=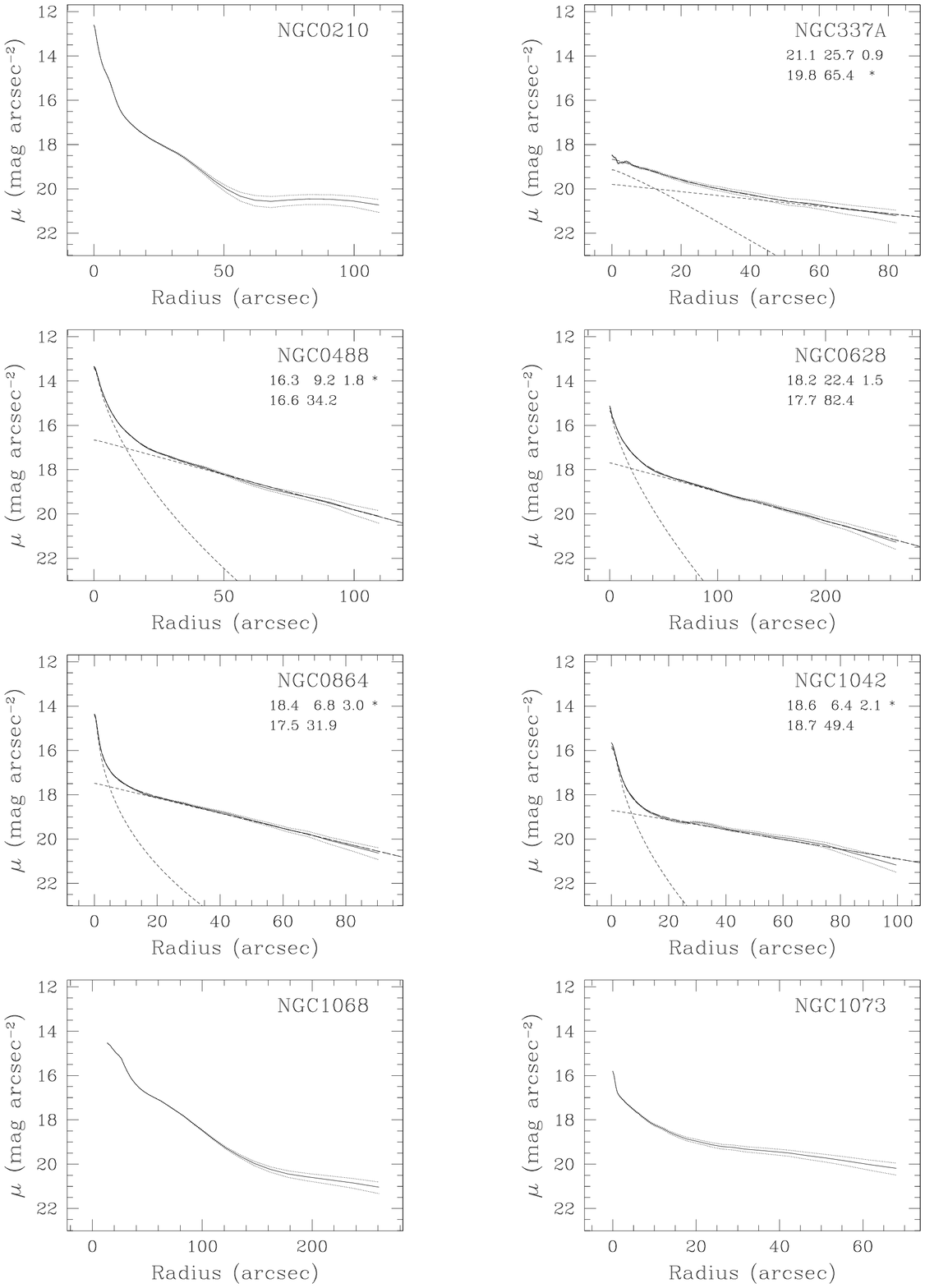,width=18cm}
\caption{Surface  brightness profiles  (solid  line) for  all  our
sample    galaxies,  with maximum   uncertainty  due to    errors in the
subtraction of sky background indicated  by the dotted lines. Bulge/disk
decompositions  are only plotted  if  at least  one component had errors
less than 30\%. Bulge and disk fits are indicated by short-dashed lines,
the sum of  the  fitted components by   a long-dashed line, and the  fit
parameters are listed below the NGC number. The bulge parameters: \mueB\
(in $K$-\magarc),
\ReB\ (in  arcsec) and \nb, respectively,  are given on  the top line,
the disk parameters:  \muo\ (in $K$-\magarc) and $h$  (in arcsec) on the
bottom line.  All   components with parameter uncertainties  larger than
30\% due to sky background   subtraction errors have   been marked by  a
star.}
\label{profilefig}
\end{figure*}

\setcounter{figure}{1}
\begin{figure*}
\psfig{figure=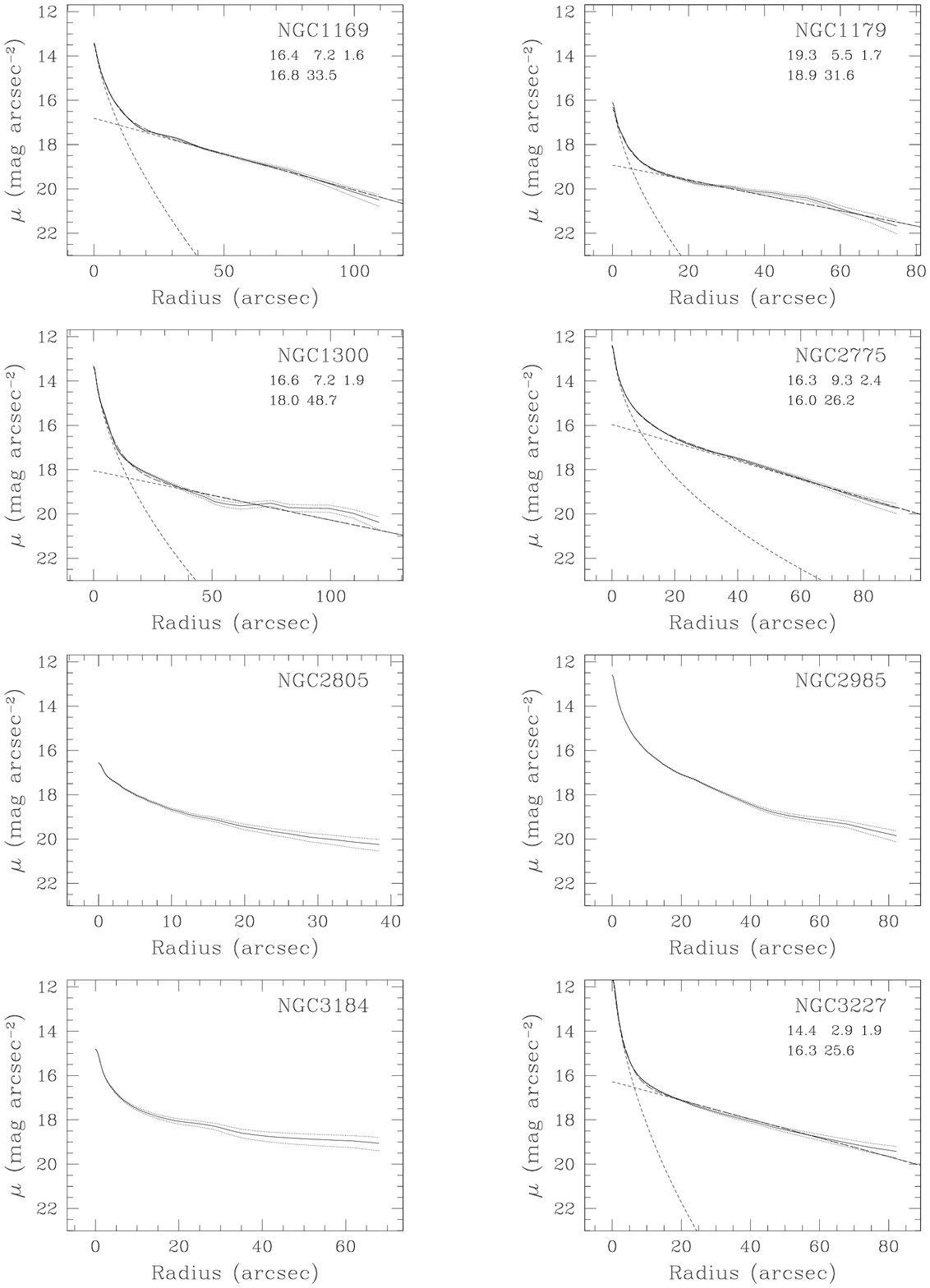,width=19cm}
\caption{Continued.}
\label{profilefig}
\end{figure*}

\setcounter{figure}{1}
\begin{figure*}
\psfig{figure=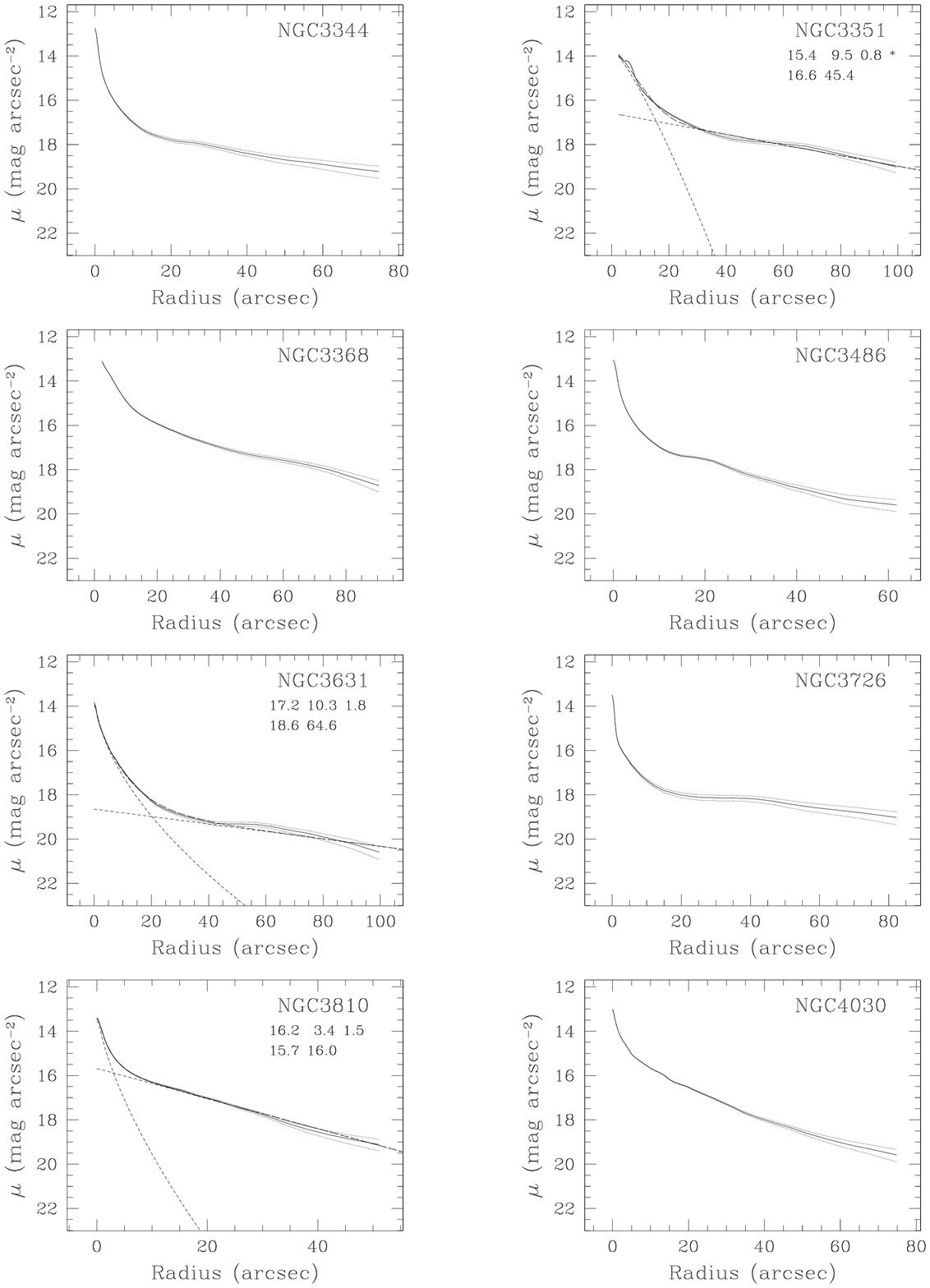,width=19cm}
\caption{Continued.}
\label{profilefig}
\end{figure*}

\setcounter{figure}{1}
\begin{figure*}
\psfig{figure=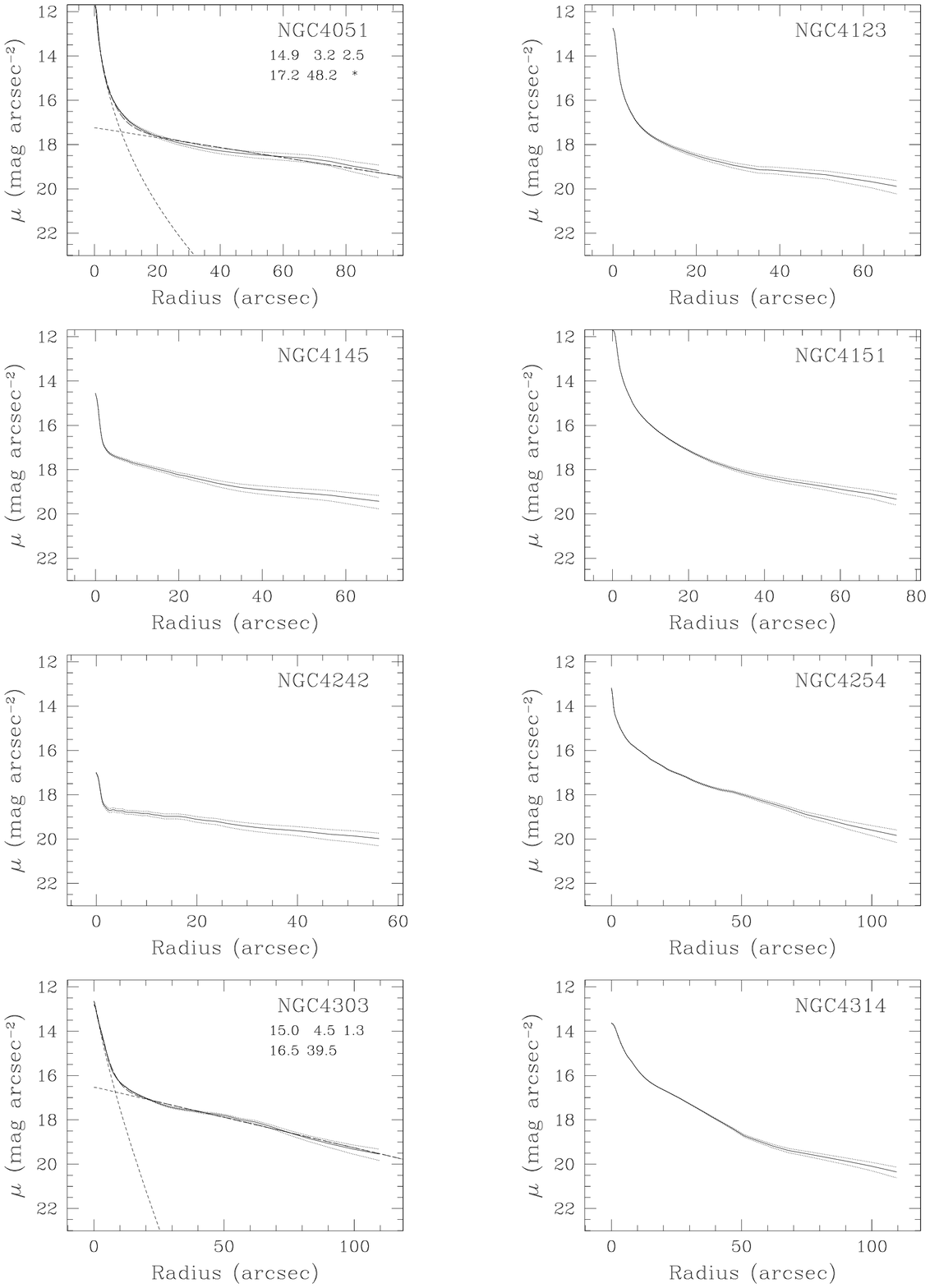,width=19cm}
\caption{Continued.}
\label{profilefig}
\end{figure*}

\setcounter{figure}{1}
\begin{figure*}
\psfig{figure=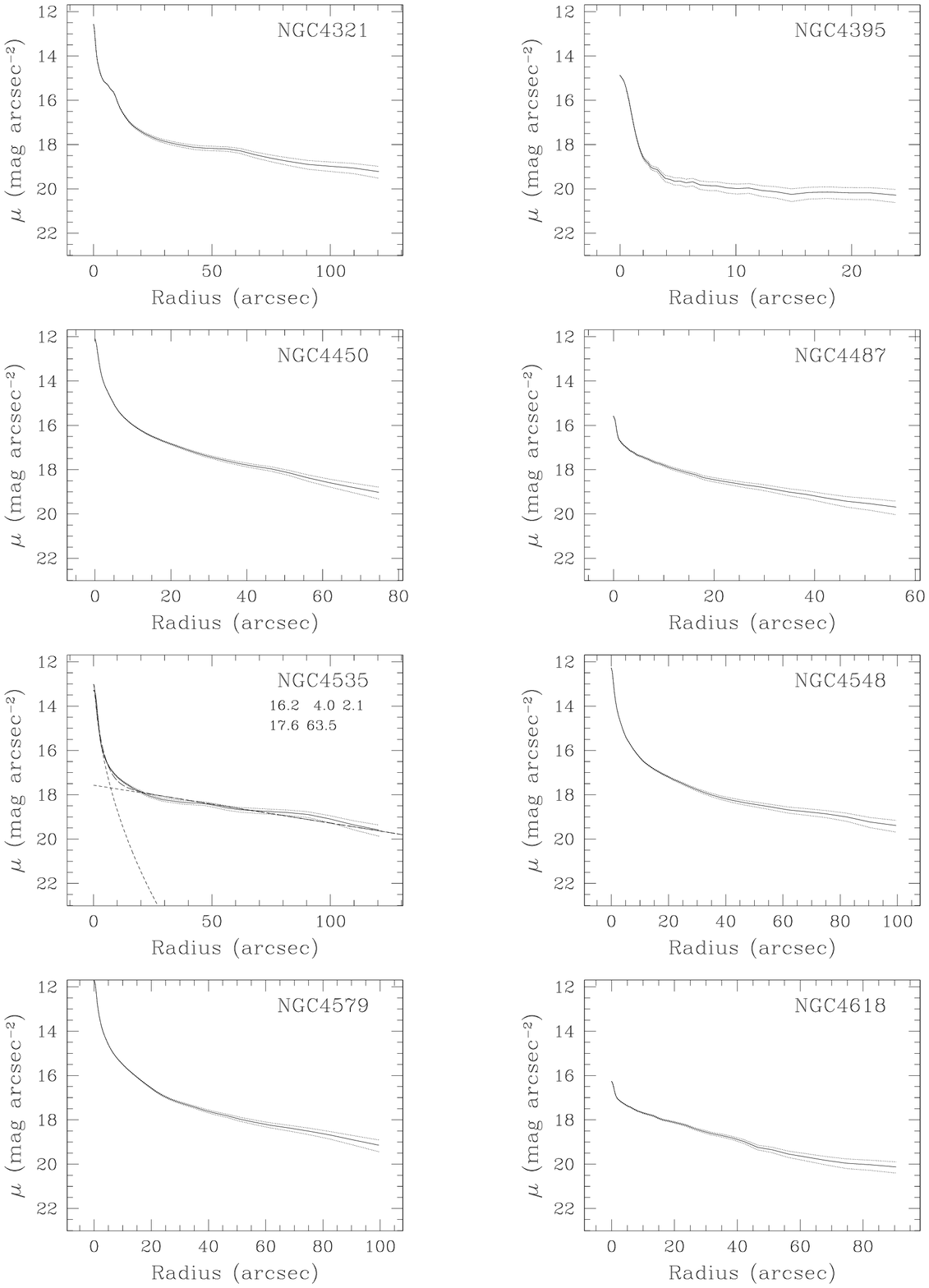,width=19cm}
\caption{Continued.}
\label{profilefig}
\end{figure*}

\setcounter{figure}{1}
\begin{figure*}
\psfig{figure=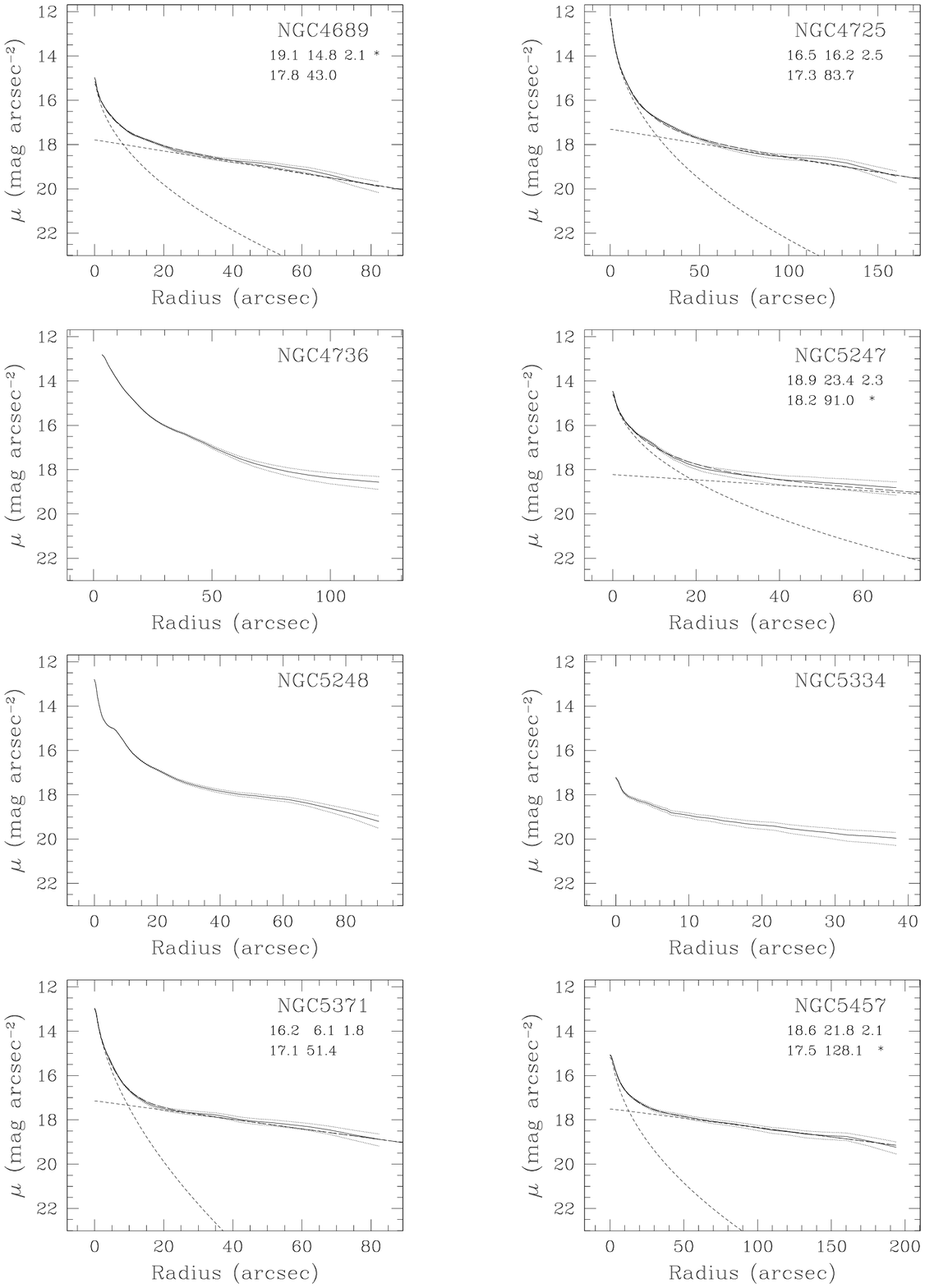,width=19cm}
\caption{Continued.}
\label{profilefig}
\end{figure*}

\setcounter{figure}{1}
\begin{figure*}
\psfig{figure=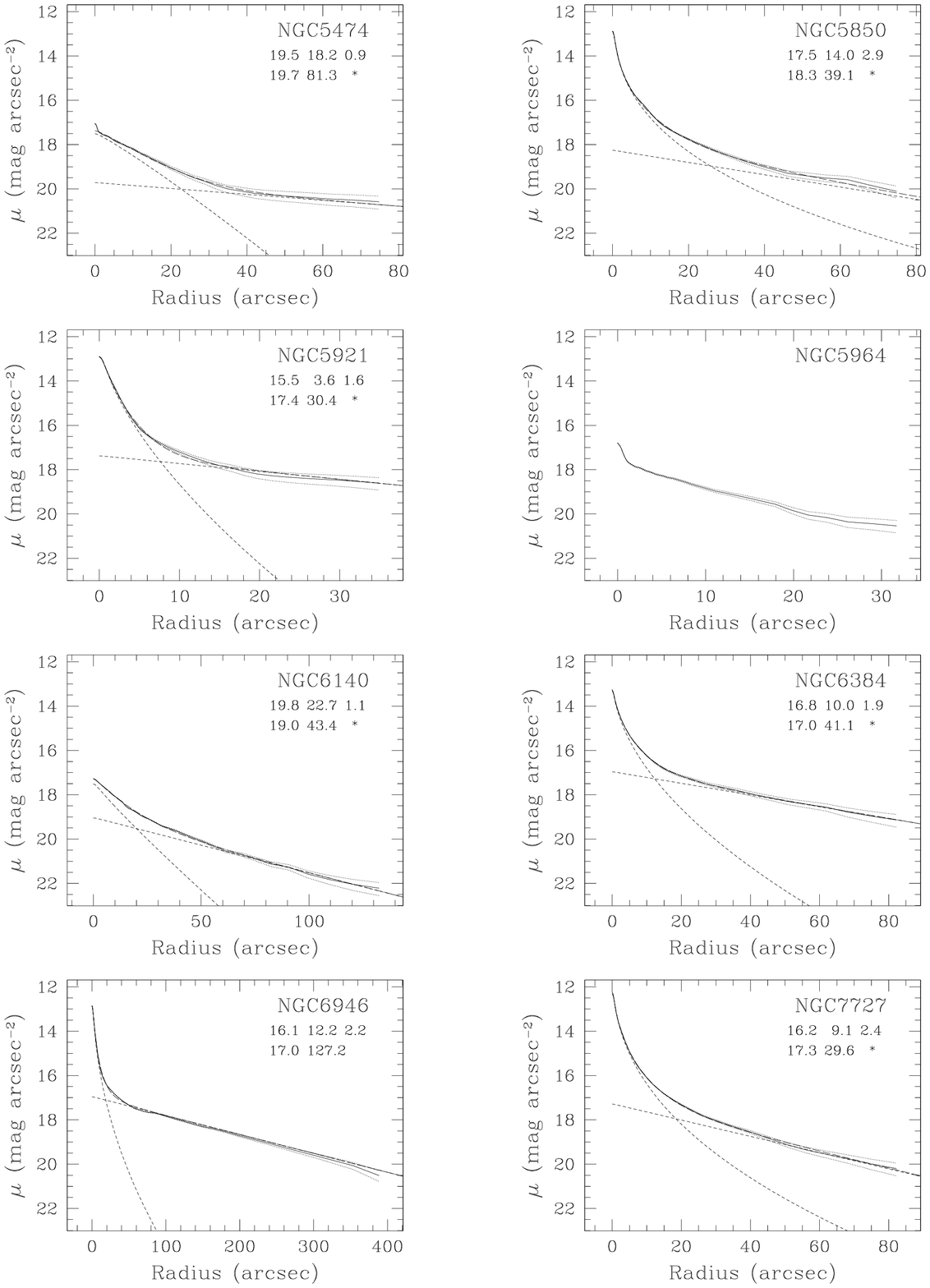,width=19cm}
\caption{Continued.}
\label{profilefig}
\end{figure*}

\label{lastpage}


\begin{thebibliography}{99}

\bibitem[Aguerri et al.~2001]{2001A&A...367..428A} Aguerri J.~A.~L., 
Balcells M., Peletier R.~F., 2001, \aap,  367, 428 

\bibitem[Allen \& Shu 1979]{1979ApJ...227...67A} Allen R.~J., Shu F.~H., 
1979, \apj,  227, 67 

\bibitem[Andredakis 1998]{1998MNRAS.295..725A} Andredakis Y.~C., 1998, 
\mnras,  295, 725 

\bibitem[Andredakis et al.~1995]{1995MNRAS.275..874A} Andredakis Y.~C., 
Peletier R.~F., Balcells M., 1995, \mnras,  275, 874 

\bibitem[Blanton et al.~2001]{2001AJ....121.2358B} Blanton M.~R., Dalcanton 
J., Eisenstein D., et al., 2001, \aj,  121, 2358 

\bibitem[\protect\citename{Block} 2001]{2001A&A...375..761B} Block D.~L., 
Puerari I., Knapen J.~H., Elmegreen B.~G., Buta R., Stedman S., Elmegreen 
D.~M., 2001, A\&A,  375, 761 

\bibitem[Byun \& Freeman 1995]{1995ApJ...448..563B} Byun Y.~I., Freeman 
K.~C., 1995, \apj,  448, 563

\bibitem[Courteau et al.~1996]{1996ApJ...457L..73C} Courteau S., de Jong 
R.~S., Broeils A.~H., 1996, \apjl,  457, L73 

\bibitem[Cross \& Driver 2002]{2002MNRAS.329..579C} Cross N., Driver S.~P., 
2002, \mnras,  329, 579 

\bibitem[de Jong 1996a]{1996A&AS..118..557D} de Jong R.~S., 1996a, \aaps,  
118, 557 

\bibitem[de Jong 1996b]{1996A&A...313...45D} de Jong R.~S., 1996b, \aap,  
313, 45 

\bibitem[de Jong \& Lacey 2000]{2000ApJ...545..781D} de Jong R.~S., Lacey 
C., 2000, \apj,  545, 781 

\bibitem[de Vaucouleurs 1948]{1948AnAp...11..247D} de Vaucouleurs G., 1948, 
Annales d'Astrophysique,  11, 247

\bibitem[de Vaucouleurs, de Vaucouleurs, \&
Corwin(1976)]{1976RC2...C......0D} de Vaucouleurs, G., de Vaucouleurs,
A., \& Corwin, J.~R.\ 1976, Second reference catalogue of bright
galaxies, 1976, Austin: University of Texas Press (RC2)

\bibitem[de Vaucouleurs et al.~1991]{1991RC3...C......0D} de Vaucouleurs 
G., de Vaucouleurs A., Corwin J.~R., Buta R.~J., Paturel G., Fouque P., 
1991, Third reference catalogue of Bright galaxies, 1991, New York : 
Springer-Verlag (RC3)

\bibitem[Disney \& Phillipps 1983]{1983MNRAS.205.1253D} Disney M., 
Phillipps S., 1983, \mnras,  205, 1253 

\bibitem[Doyon et al.~1998]{1998SPIE.3354..760D} Doyon R., Nadeau D., 
Vallee P., Starr B.~M., Cuillandre J.~C., Beuzit J., Beigbeder F., 
Brau-Nogue S., 1998, Proc. SPIE,  3354, 760

\bibitem[\protect\citename{Elmegreen} 1987]{1987ApJ...314....3E}
Elmegreen D.~M., Elmegreen B.~G., 1987, ApJ, 314, 3

\bibitem[Eskridge et al.~2000]{2000AJ....119..536E} Eskridge P.~B., Frogel 
J.~A., Pogge R.~W., et al., 2000, AJ,  119, 536

\bibitem{}{} Eskridge P.~B., Frogel J.~A., Pogge R.~W., et al., 2002,
ApJS, 143, 73

\bibitem[Graham 2001]{2001AJ....121..820G} Graham A.~W., 2001, \aj,  121, 
820 

\bibitem[Hawarden et al.~2001]{2001MNRAS.325..563H} Hawarden T.~G., Leggett 
S.~K., Letawsky M.~B., Ballantyne D.~R., Casali M.~M., 2001, MNRAS,  325, 
563

\bibitem[Hunt et al.(1998)]{1998AJ....115.2594H} Hunt, L.~K., Mannucci,
F., Testi, L., Migliorini, S., Stanga, R.~M., Baffa, C., Lisi, F., \&
Vanzi, L.\ 1998, \aj, 115, 2594

\bibitem[\protect\citename{Jarrett} 2003]{2003AJ....125..525J} Jarrett
T.~H., Chester T., Cutri R., Schneider S.~E., Huchra J.~P., 2003, AJ,
125, 525

\bibitem[\protect\citename{Jogee} 2002]{2002ApJ...575..156J} Jogee S., 
Shlosman I., Laine S., Englmaier P., Knapen J.~H., Scoville N., Wilson 
C.~D., 2002, ApJ,  575, 156 

\bibitem[Knapen et al.~2000]{2000ApJ...529...93K} Knapen J.~H., Shlosman 
I., Peletier R.~F., 2000, ApJ,  529, 93 

\bibitem{}{} Knapen J.~H., Stedman S., Bramich D.~M., Folkes S.,
2003, submitted to MNRAS (Paper~II)

\bibitem[MacArthur et al.]{} MacArthur L.A., Courteau S., Holtzman
J.A., 2003, ApJ, 582, 689

\bibitem[McCarthy et al.~2001]{2001PASP..113..353M} McCarthy D.~W., Ge J., 
Hinz J.~L., Finn R.~A., de Jong R.~S., 2001, PASP,  113, 353 

\bibitem[M{\" o}llenhoff \& Heidt 2001]{2001A&A...368...16M} M{\" 
o}llenhoff C., Heidt J., 2001, \aap,  368, 16 

\bibitem[Moriondo et al.~1999]{1999A&AS..137..101M} Moriondo G., Baffa C., 
Casertano S., et al., 1999, \aaps,  137, 101

\bibitem{}{} Mulchaey J., Regan M., 1997, ApJ, 482, L135

\bibitem{}{} Packham C., Thompson K., Zurita, A., et al., 2003,
MNRAS, submitted

\bibitem{}{} Roche P.~F., et al., 2002, Proc Spie 4841, Instrument Design
and Performance for Optical/IR Ground-Based Telescopes, eds. M.~Iye and
A.~F.~M. Moorwood

\bibitem[Sandage \& Bedke 1994]{1994cag..book.....S} Sandage A., Bedke J., 
1994, Washington, DC: Carnegie Institution of Washington with The 
Flintridge Foundation

\bibitem[Schlegel et al.~1998]{1998ApJ...500..525S} Schlegel D.~J., 
Finkbeiner D.~P., Davis M., 1998, \apj,  500, 525 

\bibitem[Sersic 1968]{1968adga.book.....S} Sersic J.~L., 1968, Cordoba, 
Argentina: Observatorio Astronomico

\bibitem[Seigar \& James 1998]{1998MNRAS.299..672S} Seigar M.~S., James 
P.~A., 1998a, \mnras,  299, 672

\bibitem[Seigar \& James 1998]{1998MNRAS.299..685S} Seigar M.~S., James 
P.~A., 1998b, \mnras,  299, 685

\bibitem{}{} Sellwood J.A., Wilkinson A., 1993, Rep. Prog. Phys.,
56, 173 

\bibitem[Simard et al.~2002]{2002ApJS..142....1S} Simard L., Willmer 
C.~N.~A., Vogt N.~P., et al., 2002, \apjs,  142, 1 

\bibitem{}{} Skrutskie M. F., et al., 1997, in The Impact of Large Scale 
Near-IR Sky Surveys, ed. F. Garzon et al. (Dordrecht: Kluwer), 187

\bibitem[Stedman \& Knapen 2001]{2001Ap&SS.276..517S} Stedman S., Knapen 
J.~H., 2001, \apss,  276, 517 

\bibitem{}{} Tran, K.-V., Simard, L., Illingworth, Franx, M., 2003, ApJ, 
in press (astro-ph/0302292)

\bibitem[Trujillo et al.~2002]{2002MNRAS.333..510T} Trujillo I., Asensio 
Ramos A., Rubi{\~ n}o-Mart\'\i n J.~A., Graham A.~W., Aguerri J.~A.~L., 
Cepa J., Guti{\' e}rrez C.~M., 2002, \mnras,  333, 510 

\bibitem[Tully(1988)]{1988ngc..book.....T} Tully, R.~B.\ 1988, Cambridge 
and New York, Cambridge University Press

\end{thebibliography}
\end{document}